\documentclass[11pt,letterpaper]{article}
\usepackage[T1]{fontenc}
\usepackage[utf8]{inputenc}
\usepackage[margin=1in]{geometry}
\usepackage{amsmath,amssymb,amsthm}
\usepackage{graphicx}
\usepackage{booktabs}
\usepackage{algorithm,algorithmic}
\usepackage{enumitem}
\usepackage{cite}
\usepackage{authblk}
\usepackage{microtype}
\usepackage[font=small,labelfont=bf]{caption}
\usepackage[hidelinks,pdfstartview=FitH]{hyperref}
\hypersetup{
  pdftitle={Controlling the False Discovery Rate via Model-Agnostic Importance Measures},
  pdfauthor={Mengtao Wen, Haojie Ren, Changliang Zou},
  pdfkeywords={Asymptotic linearity, Cross-fitting method, Data aggregation, Multiple testing, Side information}
}
\allowdisplaybreaks
\numberwithin{equation}{section}
\theoremstyle{plain}
\newtheorem{theorem}{Theorem}

\newtheorem{assumption}{Assumption}
\newtheorem{lemma}{Lemma}
\newtheorem{proposition}{Proposition}
\theoremstyle{definition}

\newtheorem{remark}{Remark}
\newcommand{\Step}[1]{\par\medskip\noindent\textbf{#1}\quad}

\usepackage{xcolor}
\newcommand\ph{$\phantom{1}$}
\def\nindep{{\,\not\!\perp\!\!\!\perp\,}}
\def\indep{{\,\perp\!\!\!\perp\,}}
\def\T{{ \mathrm{\scriptscriptstyle T} }}
\def\E{{E}}
\def\P{\mathrm{pr}}
\def\cS{\mathcal{S}}
\def\cF{\mathcal{F}}
\def\pred{V} % predictiveness measure
\def\bR{\mathbb{R}}
\def\bH{\mathbb{H}}
\def\cA{\mathcal{A}}
\def\cN{\mathcal{N}}
\def\FDR{\mathrm{FDR}}
\def\FDP{\mathrm{FDP}}
\def\cH{\mathcal{H}}
\def\cD{\mathcal{D}}
\def\cov{\mathrm{Cov}}
\def\1{\mathbb{I}}
\def\sL{\mathcal{L}}
\def\cP{\mathcal{P}}
\def\bG{\mathbb{G}}
\def\SNR{\mathrm{SNR}}
\def\cG{\mathcal{G}}
\def\cQ{\mathcal{Q}}
\def\A{\mathrm{CA}}
\def\diag{\mathrm{diag}}
\def\B{{\mathrm{E}}}
\def\coef{{b}}
\def\cM{{\mathcal{M}}}
\def\cX{{\mathcal{X}}}
\def\cY{{\mathcal{Y}}}
\def\d{{ \mathrm{d} }}
\def\cI{{\mathcal{I}}}
\def\cJ{{\mathcal{J}}}
\def\cE{{\mathcal{E}}}

\title{Controlling the False Discovery Rate via Model-Agnostic Importance Measures}
\author[1]{Mengtao Wen}
\author[2]{Haojie Ren\thanks{Corresponding author.}}
\author[1]{Changliang Zou}
\affil[1]{School of Statistics and Data Science, Nankai University, Tianjin 300071, China}
\affil[2]{School of Mathematical Sciences, Shanghai Jiao Tong University, Shanghai 200240, China}
\affil[ ]{\href{mailto:mtwen97@gmail.com}{\texttt{mtwen97@gmail.com}},
\href{mailto:haojieren@sjtu.edu.cn}{\texttt{haojieren@sjtu.edu.cn}},
\href{mailto:nk.chlzou@gmail.com}{\texttt{nk.chlzou@gmail.com}}}

\date{}

\begin{document}
\maketitle

\begin{abstract}
  Selecting covariates while controlling the false discovery rate (FDR) is critical for interpretable scientific discovery, especially with complex machine learning models. Traditional variable selection methods often rely on specific model assumptions, while Model-X knockoffs provide a general way that leverages knowledge of the covariate distribution to produce a variable selection algorithm. 
  This study puts forward a complementary method to Model-X knockoffs, named Model-Agnostic Selection with agent-X importance (MAS), which utilizes the sample-splitting strategy and a novel variable importance measure to bypass distributional assumptions on covariates. The proposed model-agnostic agent-X importance is defined to evaluate the difference of oracle predictiveness on both the original distribution and the distribution with agent variable in population level, and can be efficiently estimated by constructing the agent variable as some constant and cross-fitting strategy. The MAS method integrates a resplit-per-covariate strategy to decouple estimator dependencies and a data aggregation technique to enhance power. Theoretical guarantees for asymptotic FDR control are established under mild conditions. Extensive simulations and real-data analyses demonstrate that MAS provides competitive performances with existing works in FDR control, computational efficiency and power.
\end{abstract}

\section{Introduction}
\label{sec:intro}

It has become the current trend to employ various types of complex machine-learning or AI approaches, such as deep neural networks~\cite{DNN,DNN4Science} and language models~\cite{GPT3,LLM4Science}, for scientific discovery.
Due to the non-transparency of those approaches, it is important to identify the relevant covariates contributing to the response of interest from a large number of explanatory
variables in order to improve the interpretability of the model and the reproducibility of the results. 

Suppose \(X = (X_1, \ldots, X_p)^\T\) is a vector of \(p\)-dimensional covariates and \(Y\) is the response. 
A covariate $X_j$ is considered to be important to $Y$ if $Y$ and $X_j$ are dependent conditional on $X_{-j}=(X_1,\ldots,X_{j-1},X_{j+1}, \ldots, X_p)^\T$, denoted by \(Y\nindep X_j\mid X_{-j}\). Otherwise, \(X_j\) is unimportant to \(Y\), denoted by \(Y\indep X_j\mid X_{-j}\). In practice, the response $Y$ depends only  on a relative small subset \(\cS_1 = \{j\in[p]: Y\nindep X_j\mid X_{-j}\}\), where \([p] = \{1, \ldots, p\}\). Thus, the goal is to develop procedures to identify important variables in \(\cS_1\) as many as possible while controlling the false discovery rate (FDR)~\cite{Benjamini1995FDR}.

Many advances in FDR control are {\it model-dependent}. 
By assuming parametric models for the response conditional on the covariates, a canonical approach is to apply Benjamini-Hochberg (BH) procedure~\cite{Benjamini1995FDR} to the p-values of coefficients. For example, \cite{Javanmard2019fdr} and \cite{Ma2020} employed BH procedure with asymptotically valid p-values obtained via debiased methods~\cite{Javanmard2014,Geer2014,Zhang2013,Ning2017} for high-dimensional linear and logistic regression models, respectively. 
Another powerful class is p-value-free methods, which has attracted many attentions in the past decade. Barber et al.~\cite{Barber2015knockoff} pioneered the elegant fixed-X ``knockoff" filter under linear regression models by creating ``knockoff" variables to mimic correlation structure in original variables to achieve accurate FDR control. 
In this vein, many researchers explored customizing knockoff filters for various models, including Gaussian graphical models~\cite{Li2021ggmknockoff}, log-contrast models~\cite{Srinivasan2021compoknockoff}, and feature screening~\cite{Liu2022screenknockoff}.
In addition, Du et al.~\cite{Du2021SDA} and Dai et al.~\cite{Dai2022ds,Dai2023} applied data splitting strategy to construct symmetric statistics and then utilized the symmetry property for obtaining an approximation to the number of false discoveries in mean, linear, and generalized linear models, respectively.  

A seminal {\it model-agnostic} approach for selecting variables is the {\it Model-X} knockoff framework~\cite{Candes2018}, which extends the applicability of fixed-X knockoffs by assuming the availability of the joint distribution of $X$. 
Along this line, recent development includes the conditional randomization tests~\cite{Candes2018}, conditional permutation tests~\cite{Berrett2019CPT}, and deep feature selection using paired-input nonlinear knockoffs~\cite{lu2018deeppink}, etc. Practically, the distribution of $X$ is unknown, and thus some works considered how to generate good knockoff features, such as the methods using deep generative models~\cite{romano2020deep} and using sequential MCMC algorithm~\cite{bates2021metropolized}. However, how well the FDR control is hinges heavily on the approximation accuracy of the estimated distribution of $X$.

To effectively control FDR in a model-agnostic perspective, this paper aims to develop a complementary method to Model-X knockoffs by utilizing model-agnostic variable importance measures.  
Recently, various model-agnostic approaches for quantifying variable importance have been proposed in the literature~\cite{Alessandro2019loco,Shah2020vim,Scheidegger2022wGCM,Lundborg2022vim,Cai2024pGMC}. We consider here one commonly-used metric, which is referred to as Leave-One-Covariate-Out (LOCO) approach~\cite{Lei2018loco,Williamson2021vim,Williamson2023vim}. It involves defining a predictiveness measure and examining the difference in predictiveness of two models, i.e., the full model and the model except one covariate.   
Let $Z=(Y, X)\sim P_Z$ and consider a class $\cF$ of prediction models that map the covariates to the response. Define \(\pred(f, P_Z)\) as the predictiveness of a function \(f: \bR^p \mapsto \bR\) under the probability measure \(P_{Z}\). 
A smaller value of \(\pred(f, P_{Z})\) indicates better predictiveness of \(f\) under \(P_Z\). An example for \(\pred\) is the mean squared prediction error (MSPE) \(\pred(f, P_{Z})=\E_{P_Z}[\{Y - f(X)\}^2]\) in regression. 
The best possible predictive function among the class \(\cF\) under \(P_{Z}\) is defined as \(f^* = {\arg\min}_{f\in\cF} \pred(f, P_{Z})\). Hence, \(\pred(f^*, P_Z)\) represents the best predictiveness of \(\cF\) under \(P_Z\). 
Then, an importance measure of \(X_j\) can be defined as the difference of the predictivenesses of \(\cF\) and its sub-class \(\cF_{-j} = \{f\in\cF: f(u) = f(v) \text{ for all } u, v\in\bR^p \text{ satisfying } u_{-j} = v_{-j}\}\) (which consists of functions unrelated to \(X_j\)), say
\begin{equation}\label{eq:psi-william}
    \varphi_j =\pred(f^*_{-j}, P_Z)-\pred(f^*, P_Z),
\end{equation}
where \(f^*_{-j} = {\arg\min}_{f\in\cF_{-j}}\pred(f, P_Z)\).
Since \(\cF_{-j}\subseteq\cF\), it follows that \(\varphi_j \geq 0\) and the equality holds only when \(f^*\in\cF_{-j}\), which implies that $X_j$ has no contribution to prediction capability. 

Consequently, testing \(Y\indep X_j\mid X_{-j}\)
boils down to testing
\begin{equation}\label{eq:Hj-psi-william}
    \bH_{0, j}: \varphi_j = 0 \quad \text{versus} \quad \bH_{1, j}: \varphi_j >0, \qquad j\in[p].
\end{equation}
Some works have studied the use of sample-splitting or cross-fitting strategy to estimate $\varphi_j$ and construct asymptotically valid tests~\cite{Williamson2021vim,Williamson2023vim,Dai2024vim,Cai2022a,Chen2023}.
An off-the-shelf idea to develop the model-agnostic selection procedure is thus to apply the BH procedure with p-values for each hypothesis in \eqref{eq:Hj-psi-william}. Unfortunately, it suffers from three main issues. First, estimating \(p\) variable importance measures $\varphi_j$' s must fit \((p+1)\) functions \(\{f^*\}\cup\{f^*_{-j}: j\in[p]\}\), which brings a heavy computational burden. 
Secondly, incurred by using the same data for estimating \(\varphi_j\)'s, the potentially strong dependence between estimators would result in distortion of FDR control. 
Lastly, the sample splitting strategy is necessary for yielding an asymptotically non-degenerate null distribution~\cite{Williamson2023vim,Dai2024vim}, but it generally results in considerable power loss.

\subsection{Our main ideas and contributions}

In this article, we develop a p-value-free framework, Model-Agnostic Selection with agent-X importance (MAS), for controlling FDR via cross-fitting strategy. The proposed method integrates a refined LOCO importance measure for efficient computation, a resplit-per-covariate strategy to reduce the strong dependence among LOCO statistics, and a symmetrized data aggregation technique to enhance power. MAS can be viewed as a model-agnostic counterpart of the sample-splitting-based procedures~\cite{Du2021SDA,Dai2022ds} and serves as an alternative to the Model-X knockoffs without the need to estimate the distribution of $X$. Three major ingredients are summarized. 

\begin{itemize}
\item[1.] \emph{Agent-X importance}. We propose a dual form of the LOCO importance measure, named by {\it agent-X} importance. Instead of leaving \(X_j\) out in \eqref{eq:psi-william}, we construct an ``agent" \(\tilde{X}_j\) of \(X_j\) and evaluate the difference of predictiveness of \(f^*\) on both the original distribution $P_Z$ and the distribution of $P_{Z^j}$ where \(Z^j = (Y, (X_{-j}, \tilde{X}_j))\). Agent-X importance can be defined as
\begin{equation}\label{eq:psi}
    \psi_j = \pred(f^*, P_{Z^j}) - \pred(f^*, P_Z).
\end{equation}
The goal of replacing \(X_j\) by \(\tilde{X}_j\) in \(Z^j\) is to deteriorate the prediction capability of the best model $f^*$.
Unlike creating knockoffs that uses distribution of $X$ \cite{Candes2018}, the construction of \(\tilde{X}_j\) can be simply as a fixed constant;
see Section~\ref{sec:theory:vim} for detailed discussions. 
% The advantage of this refined measure is that there is no need to fit \(\{f^*_{-j}: j\in[p]\}\), which greatly reduces the computational cost, especially for complex machine learning models. 
Agent-X importance avoids fitting \(\{f_{-j}^*:j\in[p]\}\), substantially reducing the computational cost, especially for complex machine learning models. 
We also show through an example in Section~\ref{sec:theory:vim} that agent-X importance can preserve a signal that vanishes for LOCO as covariate correlation increases.
The \(\psi_j\) shares a conceptual similarity with some existing importance measures. Mi et al.~\cite{Mi2021vim} and Wood et al.~\cite{Wood2024vim} proposed the permutation-based importance measures, which can be regarded as a special case of \(\psi_j\) by setting \(\tilde{X}_j\) as an independent copy of \(X_j\) and implementing \(\tilde{X}_j\) and tests by permutation in samples. However, there is no rigorous guarantee on the validity of the permutation tests. 

\item[2.] \emph{Resplit-per-covariate strategy}. We employ cross-fitting strategy to estimate agent-X importance $\psi_j$.  In particular, we develop a special sample splitting strategy where the samples in each fold will be resplit independently for each covariate so that all estimated null importance measures are approximately independent. 
This strategy could largely decouple the potentially strong dependence between estimators of importance measures and thus provide valid FDR control.

\item[3.] \emph{Data aggregation}. 
To improve power, we provide a model-agnostic construction of ranking statistics by further aggregating the evidence of important variables.  
The resulting statistics improve the signal-to-noise ratio and satisfy the marginal symmetry properties, allowing us to employ the p-value-free threshold rule~\cite{Barber2015knockoff,Dai2022ds,Du2021SDA} to make discoveries. 
\end{itemize}

The newly proposed methodology is generic and is applicable for most of commonly-used predictiveness functions. In Section~\ref{sec:ivs}, we highlight the agent-X importance measures and propose the selection procedure for FDR control in detail. Asymptotic theories for FDR control are presented in Section~\ref{sec:theory}. In Section~\ref{sec:ext}, some structure-powered constructions for power enhancement are explored, including the incorporation of prior information and the covariate-adaptive importance measures. In Section~\ref{sec:simu}, numerical experiments indicate that the proposed framework works well in a wide range of settings. 

\section{Model-Agnostic Selection with Agent-X Importance}\label{sec:ivs}

\subsection{Agent-X importance measure and its estimate}\label{sec:vim}

For ease of exposition, we consider a predictiveness function \(f\mapsto\pred(f, P_Z)\) taking positive values,
such as MSPE and classification error rate, with a smaller value indicating better predictiveness of \(f\) under \(P_Z\).  

Recall the agent-X importance measure in \eqref{eq:psi}. 
Intuitively, if we perturb the distribution of \(X_j\), such as setting \(X_j = 0\), the predictiveness would get deteriorated if \(X_j\) is important to \(Y\); otherwise, the predictiveness is unaffected. 
Thus, it can be expected from the construction \eqref{eq:psi} that \(\psi_j \geq 0\) and the equality holds when \(X_j\) is unimportant to \(Y\). 
For example, if we take \(\pred(f, P_Z) = \E_{P_Z}[\{Y - f(X_{-j}, X_j)\}^2]\) and set \(\tilde{X}_j = 0\), then \(\psi_j = \E[\{f^*(X_{-j}, X_j) - f^*(X_{-j}, 0)\}^2] \geq 0\) under some mild conditions, where \(\psi_j = 0\) when \(f^*(X) = \E(Y\mid X)\) is unrelated to \(X_j\). 
Rigorous justification on non-negativeness of \(\psi_j\) for a general \(\pred\), together with more constructions of the agent variable \(\tilde{X}_j\), can be found in Section~\ref{sec:theory:vim}. 
Then, similar to \eqref{eq:Hj-psi-william}, the concerned problem can be reframed as testing the following \(p\) hypotheses simultaneously, 
\begin{equation}\label{eq:Hj-psi}
    \bH_{0, j}: \psi_{j} = 0,\quad \text{versus} \quad \bH_{1, j}: \psi_{j} > 0, \qquad j\in[p].
\end{equation}
The main goal is to develop a procedure to output an index set of selected variables $\cA$ so that the asymptotic FDR is controlled at a pre-specified level $\alpha$, i.e., $\mathop{\lim\sup}_{n,p\to\infty} \FDR_{\cA} \leq \alpha$, where
\begin{equation}\label{eq:FDR}
    \FDR_{\cA} = \E(\FDP_{\cA}), \quad \text{with}\quad \FDP_{\cA} = \frac{|\cA\cap {\cH}_0|}{1\vee|\cA|},
\end{equation}
${\cH}_0=\{j:\psi_j=0\}$ represents the true null set, and ${\cH}_1$ denotes its complement.

To ensure stable inference and accommodate versatile machine learning prediction algorithms, we incorporate the cross-fitting technique~\cite{chernozhukov2018double,Williamson2023vim} to estimate $\psi_j$, which usually leads to central limit theorems under weaker conditions than would otherwise be necessary. Suppose that \(n\) data points \(\cD = \{Z_i=(Y_i,X_i): i\in[n]\}\) are independently drawn from the distribution \(P_{Z}\). The covariates are assumed to be standardized. 
Let \(K\geq 2\) be a fixed integer representing the number of cross-fitting folds.
First, we split \(\cD\) into \(K\) folds \(\cD_1, \ldots, \cD_K\) with size \(n/K\), and get the estimator \(\hat{f}_k\) of \(f^*\) with \(\cD\setminus\cD_k\) for each fold \(k\in[K]\). 
Then, we split \(\cD_k\) into two equal-size disjoint subsets \(\cD_{k, 0}\) and \(\cD_{k, 1}\), and calculate 
\begin{equation}\label{eq:psi-cf}
    \hat{\psi}_{j, k} = \pred(\hat{f}_k, P_{Z^j, \cD_{k, 1}}) - \pred(\hat{f}_k, P_{Z, \cD_{k, 0}}),\quad j\in[p]
\end{equation}
in the \(k\)th fold, where $P_{Z,\cD}$ is the empirical distribution functions of $Z$ on the data $\cD$. 
Note that splitting \(\cD_k\) is not needed in conventional cross-fitting technique, but is necessary here for avoiding the asymptotically degenerate null distribution of \(\hat{\psi}_j\), as discussed in~\cite{Chen2023,Williamson2023vim}. Finally, we average the estimators across folds to get 
\begin{equation}\label{eq:psi-hat}
    \hat{\psi}_j = K^{-1}\sum_{k\in[K]}\hat{\psi}_{j, k},\quad j\in[p].
\end{equation}

The following proposition shows the asymptotic linearity of the proposed estimator $\hat{\psi}_j$ in \eqref{eq:psi-hat}, which serves as a foundation for establishing the selection rule for FDR control.
Let \(\dot{\pred}(f, P; h)\) be the G\^{a}teaux derivative of \(\tilde{P}\mapsto \pred(f, \tilde{P})\) at \(P\) in the direction \(h\). 
For notational convenience, let \(\nu_{0}(Z_i) := \dot{\pred}(f^*, P_Z; \delta_{Z_i} - P_Z)\) and \(\nu_j(Z_i) := \dot{\pred}(f^*, P_{Z^j}; \delta_{Z_i^j} - P_{Z^j})\), where \(Z_{i}^j = (Y_i, X_{i, -j}, \tilde{X}_{i, j})\) and \(\delta_{Z}\) is the Dirac probability measure at \(Z\).

\begin{proposition}[Asymptotic linearity]\label{prop:asy-linear}
Suppose that Assumptions~\ref{A3}--\ref{A2} 
in the Appendix~\ref{supp:agentX} hold. Then, 
\begin{equation}\label{eq:psi-asymp-fix}
    \hat{\psi}_j - \psi_{j} = \frac{2}{n}\sum_{Z_i\in\cD_1} \nu_j(Z_i) - \frac{2}{n}\sum_{Z_i\in\cD_0} \nu_0(Z_i) + o_p(n^{-1/2}),
\end{equation}
where \(\cD_\ell = \cup_{k\in[K]} \cD_{k, \ell}\) for \(\ell\in\{0, 1\}\).
\end{proposition}

This proposition is a direct counterpart of Theorem~2 in~\cite{Williamson2023vim}.
Assumptions~\ref{A3}--\ref{A2} 
are some mild conditions on the convergence of fitting algorithm, the continuity of \(f\mapsto \pred(f, P)\), and the local uniform Hadamard differentiability of \(P\mapsto \pred(f, P)\), which are relegated to the Appendix~\ref{supp:agentX}. 

From the asymptotic linear expansion,  if \(\cD_k\) is used for estimating both \(\pred(f^*, P_{Z^j})\) and \(\pred(f^*, P_Z)\) in \eqref{eq:psi-cf}, then \(\hat{\psi}_j - \psi_j = n^{-1}\sum_{Z_{i}\in\cD}\{\nu_j(Z_i) - \nu_0(Z_i)\} + o_p(n^{-1/2})\). 
For many commonly used $V$'s, such as MSPE, 
\begin{equation}\label{eq:P1}
\nu_0(Z_i) = \nu_j(Z_i) \text{ almost surely under } \psi_j = 0,    
\end{equation}
which yields the degeneration \(\sqrt{n}(\hat{\psi}_j - \psi_j) = o_p(1)\).
This degeneration also appears when estimating the LOCO importance \(\varphi_j\)~\cite{Williamson2023vim,Dai2024vim}.
Therefore, sample splitting for \(\cD_k\) in \eqref{eq:psi-cf} is useful for avoiding this issue. 
The asymptotic linearity \eqref{eq:psi-asymp-fix} shows that $\hat{\psi}_j$ is asymptotically normal centered at \(\psi_j\), and therefore $\{\hat{\psi}_j\}_{j\in\cH_0}$ are approximately symmetric around zero while $\{\hat{\psi}_j\}_{j\in\cH_1}$ tending to be positive. 

\subsection{Resplit-per-covariate strategy}\label{sec:resplit}

Although the advantage of the symmetric property of the estimators $\hat{\psi}_j$ makes it particularly appealing to apply p-value-free methods to calculate the selection threshold, we notice that there exists potentially strong dependence among these estimators, which can produce inflated FDR~\cite{Dai2022ds,Du2021SDA}. 
In particular, for any two null importances \(\psi_{j} = \psi_{\ell} = 0\), the asymptotic expressions of \(\hat{\psi}_{j}\) and \(\hat{\psi}_{\ell}\) have the same major terms when (\ref{eq:P1}) holds, and consequently \(\sqrt{n}(\hat{\psi}_{j} - \hat{\psi}_{\ell}) = o_p(1)\). 
This implies that those null importance estimators are almost the same, and it is difficult to distinguish them. 

To relieve the strong dependence, we propose the {\it resplit-per-covariate strategy}, which \emph{independently resplits} \(\cD_k\) into two equal-size disjoint subsets \(\cD_{k, 0}^j\) and \(\cD_{k, 1}^j\) for each covariate \(j\in[p]\), and replaces the estimator in \eqref{eq:psi-cf} by the following one, 
\begin{equation}\label{eq:psi-cf-R}
    \hat{\psi}_{j, k} = \pred(\hat{f}_k, P_{Z^j, \cD_{k, 1}^j}) - \pred(\hat{f}_k, P_{Z, \cD_{k, 0}^j}), \quad j\in[p].
\end{equation}
In other words, the resplit-per-covariate strategy randomizes the sample \(\cD_{0}^j = \cup_{k\in[K]} \cD_{k, 0}^j\) and \(\cD_{1}^j = \cup_{k\in[K]} \cD_{k, 1}^j\) for different covariates, which brings great benefit for relieving the dependence between null importance estimators. 
It can be shown that for \(\psi_j = \psi_\ell = 0\), 
\[
    \cov\left(\sqrt{n}\hat{\psi}_j, \sqrt{n}\hat{\psi}_\ell\right) \approx 0,
\]
under mild conditions. Rigorous justification will be provided in Section~\ref{sec:theory:random}. This implies that the resplit-per-covariate strategy is quite exquisite and powerful in terms of relieving dependence.
Moreover, this strategy does not affect the asymptotic results of individual estimator \(\hat{\psi}_j\), and accordingly, the p-value-free threshold rule~\cite{Barber2015knockoff,Dai2022ds,Du2021SDA} can be directly applicable.

\subsection{Data aggregation}\label{sec:aggreate}

We notice that the sample splitting in \eqref{eq:psi-cf-R} leads to power loss. Motivated by data aggregation strategy in \cite{Du2021SDA}, we suggest the following refinement
\begin{equation}\label{eq:psiE}
    \hat{\psi}_j^\B = \hat{\beta}_j \times \hat{\psi}_j,\qquad j\in[p],
\end{equation}
where \(\hat{\beta}_j\) is a positive auxiliary statistic converging to some quantity \(\beta_j>0\), and the superscript E is the abbreviation of \emph{enhancement}. 
The \(\hat{\psi}_j^\B\) reduces to the trivial case \eqref{eq:psi-hat} when specifying \(\hat{\beta}_j  = 1\) for all \(j\in[p]\). Consider the following decomposition,
\[
\sqrt{n}\hat{\psi}_j^\B = \sqrt{n}\hat{\beta}_j\psi_j + \sqrt{n}\beta_j(\hat{\psi}_j-\psi_j) + \sqrt{n}(\hat{\beta}_j - \beta_j)(\hat{\psi}_j - \psi_j).
\]
According to Proposition~\ref{prop:asy-linear}, the second term is asymptotically symmetric around zero, and the last term vanishes as long as \(\max_{j\in[p]}|\hat{\beta}_j - \beta_j| = o_p(1)\).  Thus, the p-value-free threshold rule is still applicable to \(\hat{\psi}_j^\B\)'s. 
Moreover, compared with \(\hat{\psi}_j\), which is decomposed as \(\sqrt{n}\hat{\psi}_j = \sqrt{n}{\psi}_j + \sqrt{n}(\hat{\psi}_j - \psi_j)\), if \(\hat{\beta}_j\) for $j\in \cH_1$ is larger than that for $j\in \cH_0$, we can expect that using \(\hat{\psi}_j^\B\) could produce a higher signal-to-noise ratio (SNR) than using \(\hat{\psi}_j\) directly. A formal justification will be detailed in Section 3.4.
Therefore, we determine the discovery set as \(\cA = \{j\in[p]: \hat{\psi}_j^\B \geq L\}\), where 
\begin{equation}\label{eq:thrsh}
    L = \inf\left\{t > 0: \frac{\sum_{j\in[p]}\1\{\hat{\psi}_j^\B \leq -t\}}{1\vee\sum_{j\in[p]}\1\{\hat{\psi}_j^\B \geq t\}} \leq \alpha\right\},
\end{equation}
and \(\alpha\in(0, 1)\) is the nominal FDR level.

A natural construction of $\hat{\beta}_j$ is to associate with the predictiveness function \(\pred\). 
By the positiveness of \(\pred\),  \(\beta_j = \pred(f^*, P_{Z^j})\) (i.e., the first term in the definition of \(\psi_j\) \eqref{eq:psi}) appears to be a potential candidate, because it satisfies that \(\beta_j = \pred(f^*, P_Z) > 0\) if \(\psi_j = 0\) and \(\beta_j > \pred(f^*, P_Z)\) if \(\psi_j > 0\). 
For the implementation, we take per-fold data to estimate $\beta_j$ in the following way,
\begin{equation}\label{eq:betaj}
    \hat{\beta}_j = K^{-1}\sum_{k\in[K]}\pred(\hat{f}_k, P_{Z^j, \cD_{k}}).
\end{equation}
It is easy to verify that $\hat{\beta}_j$ converges to $\beta_j$ in probability, so the intuitive requirement on \(\hat{\beta}_j\) is satisfied by \eqref{eq:betaj}. 
Adopting the whole fold \(\cD_k\) in \eqref{eq:betaj} instead of $\cD_{k,0}^j$ or $\cD_{k,1}^j$ could yield a more efficient estimate of \(\beta_j\) and thus deliver more stable performance in data aggregation.
In the case that we have prior knowledge about the covariate structure, it is possible to construct $\beta_j$ in a more efficient manner, which is discussed in Section \ref{sec:ext:prior}.

\subsection{MAS algorithm}\label{sec:algorithm}

We name the proposed procedure as \emph{Model-Agnostic Selection with agent-X importance} (MAS).
The complete procedure is summarized in Algorithm~\ref{alg:mas}.

\begin{algorithm}[htbp]
\footnotesize
\caption{\quad Model-Agnostic Selection with agent-X importance (MAS)}
\label{alg:mas}
\begin{algorithmic}[1]
    \REQUIRE Observed data \(\cD = \{Z_i: i\in[n]\}\), predictiveness measure function \(\pred\), construction of the agent variable \(\tilde{X}_j\), learning algorithm \(\sL\), number of folds of cross-fitting \(K\), FDR nominal level \(\alpha\);
    \STATE Split \(\cD\) into \(K\) disjoint subsets \(\cD_{1}\cup\cdots\cup\cD_{K}\) with nearly equal sizes;
    \FOR{\(k = 1, \ldots, K\)}
      \STATE Get the estimator \(\hat{f}_{k}\) with \(\cD\setminus\cD_{k}\) via learning algorithm \(\sL\);
      \FOR{\(j = 1, \ldots, p\)}
        \STATE Randomly split \(\cD_{k}\) into two nearly equal-size subsets \(\cD_{k, 0}^j\) and \(\cD_{k, 1}^j\);
        \STATE Calculate \(\hat{\psi}_{j, k} = \pred(\hat{f}_{k}, P_{Z^j, \cD_{k, 1}^j}) - \pred(\hat{f}_{k}, P_{Z, \cD_{k, 0}^j})\);
        \STATE Calculate \(\hat{\beta}_{j, k} = \pred(\hat{f}_k, P_{Z^j, \cD_{k}})\);
      \ENDFOR
    \ENDFOR
    \FOR{\(j = 1, \ldots, p\)}
      \STATE Take average: \(\hat{\psi}_{j} = K^{-1}\sum_{k \in[K]} \hat{\psi}_{j, k}\) and \(\hat{\beta}_{j} = K^{-1}\sum_{k\in[K]} \hat{\beta}_{j, k}\);
      \STATE Construct the statistic \(\hat{\psi}_j^\B = \hat{\beta}_j \times \hat{\psi}_j\);
    \ENDFOR
    \STATE Apply the threshold rule \eqref{eq:thrsh} to \((\hat{\psi}_{1}^\B, \ldots, \hat{\psi}_{p}^\B)\) and get \(L\);
    \RETURN The discovered variable set \({\cA} = \{j\in[p]: \hat{\psi}_{j}^\B \geq L\}\).
\end{algorithmic}
\end{algorithm}
 
It is clearly noted that estimating \(p\) agent-X importance measures only needs to fit \(f^*\) $K$ times, greatly lightening the computational burden, especially for complex machine learning algorithms.
In contrast, estimating \(p\) LOCO importance measures~\cite{Williamson2023vim,Chen2023} requires fitting at least \((p+1)\) functions, which is computationally prohibitive especially when \(p\) is large.
For Model-X knockoffs, it needs to fit a model with \(2p\) variables (observed covariates and their knockoffs).
Thus, MAS and Model-X knockoffs share computational complexity of the same order of \(n\) and \(p\) in model fitting, outperforming the LOCO importance measures.

\section{Theoretical Results}\label{sec:theory}

\subsection{FDR control}\label{sec:theory:fdr}

In this section, we study the theoretical properties of the proposed MAS procedure. 
Let \(p_1 = |\cH_1|\) and \(p_0 = |\cH_0|\), and define \(\sigma_j^2 = [\E\{\nu_j(Z)\}^2 + \E\{\nu_0(Z)\}^2]/2\) for \(j\in[p]\). 
We need some technical conditions. 

\begin{assumption}[Agent-X importance]\label{asmp:vim}
    (a) \(\psi_j \geq 0\); (b) the asymptotic linearity \(\sqrt{n}(\hat{\psi}_j - \psi_j) = 2{n}^{-1/2} \{\sum_{Z_i\in\cD_1^j} \nu_j(Z_i) - \sum_{Z_i\in \cD_0^j} \nu_0(Z_i)\} + R_{n, j}\) holds for each \(j\in[p]\), where \(\max_{j\in[p]} |R_{n, j}| = O_p(\vartheta_{p})\) with \(\vartheta_{p}\sqrt{\log p} \rightarrow 0\) as \(n, p\rightarrow\infty\). 
\end{assumption}

Assumption~\ref{asmp:vim}(a) imposes the non-negativeness of \(\psi_j\), which will be discussed in Section~\ref{sec:theory:vim} in details. 
Assumption~\ref{asmp:vim}(b) requires the asymptotic linearity of \(\hat{\psi}_j\), which has been shown to be satisfied in Proposition~\ref{prop:asy-linear} under some general conditions on \(\pred\).

\begin{assumption}[Moments]\label{asmp:moment}
  (a) There exists some constant \(C_{\sigma}\) such that \(0 < 1/C_{\sigma} \leq \sigma_j \leq C_{\sigma} < \infty\) for \(j\in[p]\); 
  (b) \(\E\left(\max_{j\in\{0\}\cup[p]}|{\nu}_j(Z)|^\theta\right) < \infty\) for some $\theta>4$.   
\end{assumption}

\begin{assumption}[Dependence]\label{asmp:cov}
  For any \(j\neq \ell\in\cH_0\), \(|\cov\{\nu_j(Z) - \nu_0(Z), \nu_\ell(Z) - \nu_0(Z)\}|\linebreak \leq C(\log p_0)^{-\gamma}\) holds for some constants \(C>0\) and \(\gamma > 2\).
\end{assumption}

\begin{assumption}[Signal]\label{asmp:Ssignal}
  Let \(\cH_{1*} = \{j\in[p]: (\psi_{j}/\sigma_j)\sqrt{n/\log p} \rightarrow \infty\}\) and \(p_{1*} = |\cH_{1*}|\) satisfying \(p_{1*}\rightarrow\infty\) and 
  \(p_0^{-\eta}(\log p_0)^{-1/2}[\{p_{1*}(\log p_0)\}\vee p_1] \rightarrow 0\) for some constant \(\eta>0\).
\end{assumption}

Assumption~\ref{asmp:moment} imposes some moment conditions on G\^{a}teaux derivatives.
The covariance between null importance estimators is characterized by Assumption~\ref{asmp:cov}; for many commonly used \(\pred\)'s for which (\ref{eq:P1}) holds, such as MSPE, this covariance becomes zero and thus Assumption~\ref{asmp:cov} naturally holds.
Assumption~\ref{asmp:Ssignal} requires that the number of strong signals should diverge as $p\to\infty$. Similar conditions to Assumptions~\ref{asmp:moment}--\ref{asmp:Ssignal} also appeared in \cite{Dai2023,Du2021SDA}.

\begin{assumption}[Auxiliary statistic]\label{asmp:beta}
    (a) There exist \(\{\beta_j: j\in[p]\}\) such that \(0 < 1/C_{\beta} \leq \beta_j \leq C_{\beta}\linebreak < \infty\) for some constant \(C_{\beta}\); (b) \(\max_{j\in[p]}|\hat{\beta}_{j} - \beta_{j}| = O_p(\kappa_p)\) with \(\kappa_p (\log p) \rightarrow 0\) as \(n, p\rightarrow\infty\).
\end{assumption}

Assumption~\ref{asmp:beta} summarizes general requirements on \(\beta_j\) and \(\hat{\beta}_j\), which need \(\beta_j\)  bounded and \(\hat{\beta}_j\) converging to \(\beta_j\) uniformly. This will be verified for the construction \eqref{eq:betaj} in Section~\ref{sec:theory:beta}. 

\begin{theorem}\label{THM:FDR}
    Suppose that Assumptions~\ref{asmp:vim}--\ref{asmp:beta} hold. If \(p = O(n^r)\) for some constant \(r > 0\), then for any $\alpha\in (0,1)$, the MAS procedure in Algorithm~\ref{alg:mas} has \(\FDP_{\cA} \leq \alpha + o_p(1)\) as \(n, p\rightarrow\infty\). It follows that \({\lim\sup}_{n, p \rightarrow\infty} \FDR_{\cA} \leq \alpha\).
\end{theorem}

Theorem \ref{THM:FDR} implies that the MAS procedure can control the FDR level asymptotically in high-dimensional settings. 
Note that except for the \(\hat{\beta}_j\) in Algorithm~\ref{alg:mas}, other constructions of \(\hat{\beta}_j\) are also suitable for Theorem~\ref{THM:FDR}, as long as Assumption~\ref{asmp:beta} is satisfied. 
The MAS procedure can be applied to various types of data by specifying a proper \(\pred\) as displayed in \ref{M1}--\ref{M3} in the next section. 

\begin{remark}
    The FDR control of the MAS procedure primarily relies on the asymptotic normality of \(\hat{\psi}_j\), which further depends on the convergence of \(\hat{f}_k\) to \(f^*\). 
    In contrast, the Model-X knockoff framework requires exact knowledge of the joint distribution of \(X\) to generate valid knockoff variables~\cite{Candes2018}. When this distribution is unknown, approximate methods may be employed~\cite{romano2020deep}; however, the validity of FDR control then hinges on the quality of the approximation. 
    Meanwhile, Dai et al.~\cite{Dai2022ds,Dai2023} and Du et al.~\cite{Du2021SDA} also require the asymptotic normality to achieve the asymptotic symmetry of the statistic to enable FDR control, but they assume parametric models, different from the model-agnostic methods here.
\end{remark}

\subsection{Non-negativeness of agent-X importance measure}\label{sec:theory:vim}

Now, we first show the non-negativeness of \(\psi_j\) with mild conditions and then provide some concrete examples of \(\pred\) and constructions of \(\tilde{X}_j\). Let 
\(\cP_{0, j} = \{P_{\{Y, (X_{-j}, W)\}}: W\) is some random variable and \(W\indep Y\mid X_{-j}\}\) 
be the set of probability measures satisfying conditional independence on the \(j\)th element. 

\begin{proposition}\label{prop:psi0}
    Assume that \(P\mapsto {\pred}(f^*, P)\) is a linear functional and satisfies the following conditions for any \(P_0\in\cP_{0, j}\): (a) $\dot{\pred}(f^*, P_Z; P_0 - P_Z) = 0$ if \(P_Z\in\cP_{0, j}\); and (b) $\dot{\pred}(f^*, P_Z; P_0 - P_Z) \geq 0$ if \(P_Z \not\in\cP_{0, j}\).
    If the agent variable \(\tilde{X}_j\) is constructed such that \(P_{\{Y, (X_{-j}, \tilde{X}_j)\}} \in\cP_{0, j}\), then \(\psi_j \geq 0\) and the equality holds when \(X_j\indep Y \mid X_{-j}\).
\end{proposition}

The conditions on \({\pred}\) in Proposition~\ref{prop:psi0} define the properness of predictiveness in the agent-X importance measure.
The justification of the validity of (a)--(b) depends on the concrete form of \(\pred\). We consider the following popular choices, which are suitable for different types of responses:
\begin{enumerate}[label={(M\arabic*)}]
  \itemindent15pt
	  \item \(\pred(f, P_Z) = \E[\{Y - f(X)\}^2]\) in regression.\label{M1}
    \item \(\pred(f, P_Z) = -\E[Y\log\{f(X)\} + (1-Y)\log\{1-f(X)\}]\) for binary response.\label{M2}
    \item \(\pred(f, P_Z) = -\E[Y\log\{f(X)\} - f(X) - \log(Y!)]\) for non-negative integer response.\label{M3}
\end{enumerate}
We show that the conditions (a)--(b) in Proposition~\ref{prop:psi0} hold for the above three predictiveness functions in Appendix~\ref{supp:agentX}. Other constructions of proper \(\pred\) are worth future exploration.

Proposition~\ref{prop:psi0} indicates that the construction of \(\tilde{X}_j\) should satisfy \(\tilde{X}_j\indep Y\mid X_{-j}\), which can be easily fulfilled by some constructions without seeing \(Y\), such as setting \(\tilde{X}_j\) as a constant or an independent copy of \(X_j\) (which can be approximately implemented in practice by permuting observations of \(X_j\)).
Different choices of \(\tilde{X}_j\) entail a trade-off between the magnitude of \(\psi_j\) under alternative hypotheses and the finite-sample approximation to the asymptotic normality of the estimator \(\hat{\psi}_j\).
To illustrate, we take MSPE in~\ref{M1} as the predictiveness function and consider linear models, i.e., \(f^*(x) = \E(Y\mid X = x) = x^\T {\coef}^*\) for some \({\coef}^*\in\bR^p\).
It can then be shown that 
\[
\psi_j = \E\{f^*(X_{-j}, X_j) - f^*(X_{-j}, \tilde{X}_j)\}^2 = ({\coef}^*_j)^2 \E(X_j - \tilde{X}_j)^2,
\]
where \({\coef}_j^*\) is the \(j\)th component of \({\coef}^*\). 
For a standardized \(X_j\), \(\psi_j = (1+c^2)({\coef}_j^*)^2\) if \(\tilde{X}_j = c\) for some constant \(c\), and \(\psi_j = 2({\coef}_j^*)^2\) if \(\tilde{X}_j\) is an independent copy of \(X_j\).
Consequently, setting \(\tilde{X}_j = c\) for a larger positive \(c\) yields a larger \(\psi_j\) under alternative hypotheses, and using an independent copy of \(X_j\) (approximated by permuting observations of \(X_j\)) as \(\tilde{X}_j\) is equivalent to setting \(\tilde{X}_j = 1\). 
Intuitively, larger \(\psi_j\) under alternative hypotheses would increase detection power in multiple testing, so it is natural to set \(\tilde{X}_j\) as a large positive constant \(c\). 
However, residual terms (such as \(\dot{\pred}(\hat{f}_k, P_{Z^j}; h_{n, k}) - \dot{\pred}(f^*, P_{Z^j}; h_{n, k})\) with \(h_{n, k} = \{n/(2K)\}^{1/2}(P_{Z^j, \cD_{k, 1}^j} - P_{Z^j})\)) that vanish asymptotically in the expansion of \(\hat{\psi}_j\) can become non-negligible for large \(c\) in finite samples, because the fitting error of \(\hat{f}_k\) is amplified at this time. 
This can degrade the approximate normality of \(\hat{\psi}_j\) and, in turn, affect the FDR control of the MAS procedure in practice. 
Therefore, this trade-off needs to be balanced when choosing \(\tilde{X}_j\), which is numerically examined in Section~\ref{sec:simu:agentX}. 
For standardized covariates, we recommend setting \(\tilde{X}_j = z_{0.95}\) (where \(z_q\) denotes the \(q\)th standard normal quantile) as a practical compromise. 

\begin{remark}
\label{rem:correlated-covariates}
Agent-X importance can preserve a signal that vanishes for LOCO as covariate correlation increases. 
Consider the linear regression model \(Y = b_1^*X_1 + b_2^*X_2 + \varepsilon\), where \(\E(\varepsilon \mid X_1,X_2) = 0\), \(\E(\varepsilon^2) < \infty\), and \((X_1,X_2)\) is jointly Gaussian with zero means, unit variances, and correlation \(\rho \in (-1,1)\).
Under MSPE in \ref{M1}, the LOCO importance and the agent-X importance with \(\tilde{X}_j = c\) satisfy \(\varphi_j = (b_j^*)^2(1-\rho^2)\) and \(\psi_j = (b_j^*)^2(1+c^2)\) for \(j=1,2\).
Consequently, for fixed \(b_j^* \neq 0\), the LOCO importance approaches zero as \(|\rho|\) approaches one, while the agent-X importance remains unchanged and strictly positive.
\end{remark}

\begin{remark}
  Notice that our hypothesis testing problem $\bH_{0,j}$ in (\ref{eq:Hj-psi}) is not exactly equivalent to the original conditionally independent testing problem. However, by Proposition~\ref{prop:psi0}, 
  we know that \(X_j \indep Y\mid X_{-j}\) implies $\psi_j=0$ and therefore $\cS_0 = \{j\in[p]: X_j\indep Y\mid X_{-j}\}\subseteq\cH_0$.
  Accordingly, the result in Theorem 1 is still valid for the modified FDR notion with $\cS_0$ replacing $\cH_0$ in \eqref{eq:FDR}.
\end{remark}

\subsection{Analysis of resplit-per-covariate strategy}\label{sec:theory:random}

Next, we demonstrate the working mechanism of the resplit-per-covariate strategy. To rigorously state our results, we introduce \(\gamma_{j, i} = \1(Z_i \in \cD_1^j)\).  
Then, we rewrite the asymptotic linearity in Assumption~\ref{asmp:vim} as
\begin{equation}\label{eq:psi-eps}
    \frac{\sqrt{n}}{2}(\hat{\psi}_j - \psi_j) = \frac{1}{\sqrt{n}}\sum_{i\in[n]} \{\gamma_{j, i}\nu_j(Z_i) - (1-\gamma_{j, i})\nu_0(Z_i)\} + R_{n, j}.
\end{equation}
For any fixed \(j\in[p]\), the \(\gamma_{j, i}\)'s are weakly dependent as we have the constraint $\sum_{i\in[n]}\gamma_{j, i}=n/2$. The following lemma shows that we can find a set of independent Bernoulli variables \(\{\xi_{j, i}: i \in [n], j \in [p]\}\), such that the major term in \eqref{eq:psi-eps}, ${n}^{-1/2}\sum_{i\in[n]} \{\gamma_{j, i}\nu_j(Z_i) - (1-\gamma_{j, i})\nu_0(Z_i)\}$, can be approximated by its counterpart with \(\gamma_{j, i}\) replaced by \(\xi_{j, i}\), making it possible to analyze two agent-X importance estimators \(\hat{\psi}_j\) and \(\hat{\psi}_\ell\) jointly.

\begin{lemma}\label{lem:removeDep}
    Suppose that Assumption~\ref{asmp:moment} holds. 
    Then, there exists \(\{\xi_{j, i}: i \in [n], j \in [p]\}\) satisfying that \(\xi_{j, i}\) across \(i\in[n]\) and \(j\in[p]\) are independent Bernoulli random variables with probability \(1/2\), such that 
    \[\begin{split}
        &\P\left(\max_{j\in[p]}|{\bG}_{j} - \tilde{\bG}_{j}| \geq C[(\log p)^{3}/n]^{1/4}\right) \leq  3(\log n)^{-1},
    \end{split}\]
    where \(C > 4C_{\sigma}/(1-a)\) for some \(a\in(0, 1)\), and 
    \[\begin{split}
        \bG_{j} &= \frac{1}{\sqrt{n}}\sum_{i \in [n]} \left\{\gamma_{ j, i} \nu_j(Z_i) - (1-\gamma_{j, i}) \nu_0(Z_i)\right\},\\
        \tilde{\bG}_{j} &= \frac{1}{\sqrt{n}}\sum_{i \in [n]} \left\{\xi_{j, i} \nu_j(Z_i) - (1-\xi_{j, i}) \nu_0(Z_i)\right\}.
    \end{split}\] 
\end{lemma}

Lemma~\ref{lem:removeDep} implies that the major term of \eqref{eq:psi-eps} is asymptotically equivalent to \(\tilde{\bG}_{j}\), a summation of independent and identically distributed random variables.
Accordingly, if \(\nu_j(Z) = \nu_0(Z)\) almost surely for \(\psi_j = 0\), the covariance between two null importance estimators \(\hat{\psi}_{j}\) and \(\hat{\psi}_{\ell}\) is approximately zero, i.e.,
\[
    \cov(\sqrt{n}\hat{\psi}_{j}/2, \sqrt{n}\hat{\psi}_{\ell}/2) \approx \cov(\tilde{\bG}_{j}, \tilde{\bG}_{\ell}) = 0,\quad j\neq \ell,
\]
where the last equality holds due to the independence between \(\xi_{j, i}\) and \(\xi_{\ell, i}\). 
This reveals that the resplit-per-covariate strategy makes the strong dependence between null importance estimators asymptotically vanish, helping MAS perform more stable and reliable in FDR control.

\subsection{Justification of data aggregation}\label{sec:theory:beta}

Finally, we show that the construction of \(\hat{\beta}_j\) in \eqref{eq:betaj} satisfies Assumption~\ref{asmp:beta}, and justify the effect of data aggregation \eqref{eq:psiE}. 

Similar to Proposition~\ref{prop:asy-linear}, it can be derived  that 
\begin{equation}\label{eq:beta-expr}
    \sqrt{n}(\hat{\beta}_j - \beta_{j}) 
    = \frac{1}{\sqrt{n}}\sum_{i\in[n]} \nu_j(Z_i) + \tilde{R}_{n, j},
\end{equation}
where \(\tilde{R}_{n, j}\) is the corresponding residual term. For the sake of conciseness, we directly assume that the asymptotic expression \eqref{eq:beta-expr} holds. 

\begin{proposition}\label{prop:beta}
    (a) If the assumptions in Proposition~\ref{prop:psi0} hold, then \(\beta_j \geq \pred(f^*, P_Z)\). 
    (b) Suppose that Assumption~\ref{asmp:moment} and the asymptotic expression \eqref{eq:beta-expr} hold. Assume that \(\max_{j\in[p]} |\tilde{R}_{n, j}| = O_p(\varrho_p)\) for some sequence \(\varrho_p\) satisfying \(\{(\log p)^{1/2}\vee \varrho_p\}(\log p)n^{-1/2} \rightarrow 0\). 
    Then, \(\max_{j\in[p]}|\hat{\beta}_j - \beta_j| = o_p\{(\log p)^{-1}\}\).
\end{proposition}

Proposition~\ref{prop:beta}(a) shows that \(\beta_j\)'s lower bound in Assumption~\ref{asmp:beta}(a) holds. The analysis of \(\beta_j\)'s upper bound depends on the concrete form of \(\pred\), as illustrated for \ref{M1}--\ref{M3} in the Appendix~\ref{supp:proof}. 
A heuristic approach to satisfy this assumption is taking \(\pred(f, P_Z)\wedge M\) as the predictiveness measure for a large constant \(M > 0\).
Proposition~\ref{prop:beta}(b) shows that Assumption~\ref{asmp:beta}(b) holds under some mild conditions.

Now, we justify that $\hat{\psi}_j^E$ increases the SNR by aggregating $\hat{\beta}_j$. Define the SNR as
\[
    \SNR(\hat{\psi}_j^\B; \cH_1, \cH_0) = \frac{p_1^{-1}\sum_{j\in\cH_1} \E(\hat{\psi}_j^\B)^2}{p_0^{-1}\sum_{j\in\cH_0} \E(\hat{\psi}_j^\B)^2}.
\]
Note that \(\hat{\psi}_j\)'s SNR is a special case by setting \(\hat{\beta}_j = 1\) in \(\hat{\psi}_j^\B\) for all \(j\in[p]\). 
The following proposition compares the SNRs of \(\hat{\psi}_j^\B\) and \(\hat{\psi}_j\).

\begin{proposition}\label{prop:power}
    Suppose that Assumptions~\ref{asmp:vim}--\ref{asmp:moment} hold, and the asymptotic expression \eqref{eq:beta-expr} holds for \(\beta_j = \pred(f^*, P_{Z^j})\). 
    Assume that \(\max_{j\in[p]} \E(R_{n, j}^4) \vee \E(\tilde{R}_{n, j}^4) = o(1)\), \(\max_{j\in\cH_1}\psi_j \leq C\) and \(\min_{j\in\cH_1} \sqrt{n}\psi_j \rightarrow c\in(0, +\infty]\) as \(n, p\rightarrow\infty\).  
    Then,  
    \[
    \frac{\SNR(\hat{\psi}_j^\B; \cH_1, \cH_0)}{\SNR(\hat{\psi}_j; \cH_1, \cH_0)}\times\frac{\sum_{j\in\cH_1}(n\psi_j^2/4 + \sigma_j^2)\pred^2(f^*, P_Z)}{\sum_{j\in\cH_1}(n\psi_j^2/4 + \sigma_j^2)\pred^2(f^*, P_{Z^j})} \rightarrow 1.\]    
\end{proposition}

The uniform convergence of the residuals of \(\hat{\psi}_j\)'s and \(\hat{\beta}_j\)'s in the fourth moment are required in Proposition~\ref{prop:power}, which can be satisfied by convergence in probability and some additional conditions (e.g., the boundedness or the uniform integrability). 
By this result, compared to \(\hat{\psi}_j\), the SNR is improved by the construction \(\hat{\psi}_j^\B\) due to the fact that \(\beta_j^2 = \pred^2(f^*, P_{Z^j}) > \pred^2(f^*, P_Z)\) for \(j\in\cH_1\).

\section{Structure-Powered Enhancement}\label{sec:ext}

\subsection{Incorporating prior information}\label{sec:ext:prior}

In many applications, covariates satisfy some specific structures, for example, the signals often appear in groups. In the MAS framework, we can construct specific \(\beta_j\) by integrating these structures to improve power. This section introduces the construction of \(\beta_j\) under the group structure and the ordered structure, respectively. 
The estimation of \(\beta_j\) can be given similar to \eqref{eq:betaj}, so we omit the details for conciseness.

\textbf{Group Structure.} 
Suppose that \(p\) variables are partitioned into \(d\) groups \([p] = \cG_1\cup\cdots\cup\cG_d\) according to some natural structure independent of the data \(\cD\). For example, researchers would like to identify genes that have a significant impact on some disease from thousands of genes that may be affiliated with different biological pathways. Genes within the same pathway are highly functionally related, and their expression levels may show strong similarity.
In such applications, the goal is to make individual discoveries and use the group structure to increase the detection power~\cite{Hu2010gbh}. 
Intuitively, we can set \(\beta_{j} = \beta_{k}\) whenever \(j\) and \(k\) are in the same group. In particular, we can take \(\beta_{j} = \pred(f^*, P_{Z^{\cG_{\ell(j)}}})\), where \(\ell: [p] \mapsto [d]\) is a function mapping the variable index to the group index, and \(P_{Z^\cS}\) is the distribution of \(Z^\cS = (Y, X_{-\cS}, \tilde{X}_{\cS})\) for a set \(\cS\).

\textbf{Order Structure.} In an ordered setting, we may believe that the hypotheses early in the list are more likely to be true important variables than those later in the list~\cite{Li2018SABHA}. 
To incorporate such structure, we hope to construct \(\beta_{1} \geq \cdots \geq \beta_{p} > 0\). 
Consider the sequence of predictiveness measures \(\pred(f^*, P_{Z^{\cQ_j}})\), where \(\cQ_j = [p]\setminus[j]\) for \(j\in[p]\). It is worth noting that this sequence is monotonic increasing, i.e., \(\pred(f^*, P_{Z^{\cQ_1}}) \geq \cdots \geq \pred(f^*, P_{Z^{\cQ_p}})\) due to the nested property \(\cQ_1\supset\cdots\supset\cQ_p\). Thus, we can set \(\beta_{j} = \pred(f^*, P_{Z^{\cQ_j}})\), which satisfies the decreasing requirement. 

If multiple structures exist, an approach for incorporating them is to take means over various sources of \(\beta_j\). 
For example, if we have \(\beta_j^{(b)}\) from the \(b\)th structure, then a composite construction is 
\begin{equation}\label{eq:beta-combine}
    \beta_j = h\left(\beta_j^{(1)}, \ldots, \beta_j^{(B)}\right),         
\end{equation}
where  \(B\) is the number of structures and \(h: \bR_+^B \rightarrow \bR_+\) is some mean function, such as arithmetic mean, geometric mean, or harmonic mean. 
Note that the initial construction \(\beta_j = \pred(f^*, P_{Z^j})\) in Section~\ref{sec:aggreate} can also be incorporated as individual features in this way.
The estimators for \(\beta_j\) in \eqref{eq:beta-combine} can be set as \(\hat{\beta}_j = h(\hat{\beta}_j^{(1)}, \ldots, \hat{\beta}_j^{(B)})\), where each estimator \(\hat{\beta}_j^{(b)}\) is estimated respectively.
The justification of Assumption~\ref{asmp:beta} for \(\beta_j\) in \eqref{eq:beta-combine} and its estimator \(\hat{\beta}_j\) can be derived similar to Proposition~\ref{prop:beta}.

\subsection{Covariate-adaptive importance}\label{sec:ext:covariate}

For some learning algorithms possessing solution paths, such as stepwise regression and tree-based methods, some variables are selected to add or remove in the current model at each step. 
Intuitively, we can accumulate the local importance measures in steps where \(X_j\) is selected, as the covariate-adaptive importance.

In particular, a learning algorithm \(\sL\) is fitted on \(\cD\setminus\cD_k\), providing the solution path \(\{(u_{k, 1, t}, \ldots, u_{k, p, t})\in\bR^p: t\in [T]\}\) and the estimators in the solution path \(\hat{f}_k = \{\hat{f}_{k, t}: t\in[T]\}\), where \(T\) is the number of steps, and \(u_{k, j, t}\) indicates the addition (\(u_{k, j, t} = 1\)), removal (\(u_{k, j, t} = -1\)), or no action (\(u_{k, j, t} = 0\)) of \(X_j\) in the \(t\)th step, respectively. Then, \(\{X_j: \sum_{i = 1}^t u_{k, j, i} > 0\}\) are the variables in the model \(\hat{f}_{k, t}\) of the \(t\)th step.
We take the tree-based algorithms as an example to explain these notations.
A tree is grown by iteratively picking a covariate and a cutting point according to some purity criterion, so each time child nodes are generated is a step in the solution path, and the subtrees starting from the root node corresponds to the estimators \(\hat{f}_{k, t}\)'s. 
Given a predictiveness measure \(\pred\), such as the MSPE, we can consider two predictiveness locally in the \(t\)th step, \(\pred(\hat{f}_{k, t}, P_{Z^j})\) and \(\pred(\hat{f}_{k, t}, P_{Z})\), which are evaluated with independent testing data \(\cD_{k, 1}^j\) and \(\cD_{k, 0}^j\), respectively. 
Finally, summarize the local agent-X importance in steps where \(X_j\) is selected, 
\[
\hat{\psi}_{j, k}^\A = \sum_{t = 1}^T \{\pred(\hat{f}_{k, t}, P_{Z^j, \cD_{k, 1}^j}) - \pred(\hat{f}_{k, t}, P_{Z, \cD_{k, 0}^j})\}\1(u_{k, j, t} \neq 0),
\]
and take average across different folds as covariate-adaptive importance 
\begin{equation}\label{eq:psi-covariate}
\hat{\psi}_j^\A = \frac{1}{K}\sum_{k = 1}^K \hat{\psi}_{j, k}^\A.    
\end{equation}

The covariate-adaptive importance takes local models \(\{\hat{f}_{k, t}: u_{k, j, t} \neq 0\}\) as the evaluation models instead of using \(\hat{f}_k\). These importance evaluations of \(X_j\) may be more effective because \(X_j\) is often important in steps where \(X_j\) is selected.
We apply the threshold rule \eqref{eq:thrsh} to \((\hat{\psi}_1^\A, \ldots, \hat{\psi}_p^\A)\) to get \(L^\A\) and then the discovery set is \(\cA^\A = \{j\in[p]: \hat{\psi}_j^\A \geq L^\A\}\). In Section~\ref{sec:simu}, we observe that the process with this covariate-adaptive importance performs better than Algorithm \ref{alg:mas}
with higher power. Rigorous justifications on this threshold rule are left for future work.

\section{Numerical Studies}\label{sec:simu}

\subsection{Simulation settings}

In this section, we illustrate the breadth of applicability of our proposed MAS method by studying its performance on simulated data and a real-data example.

We consider three representative models, as follows: 
\begin{enumerate}[label={(S\arabic*)}]
    \itemindent10pt  
    \item \(Y = p_1^{-1/2} \sum_{j\in\cH_1} \delta X_j + \varepsilon\), where \(\varepsilon\sim\cN(0, 1)\) and \(X\sim\cN(0, (0.6^{|i-j|})_{p\times p})\).\label{S1}
    \item \(Y = p_1^{-1/2} \sum_{j\in\cH_1} \delta(X_j + X_j^2)/2 + \varepsilon\), where \(\varepsilon\sim \cN(0, 1)\) and \(X \sim\cN(0, (0.6^{|i-j|})_{p\times p})\).\label{S2}
    \item \(Y = 2X_{\cH_{1, 1}}X_{\cH_{1, 2}}\1(X_{\cH_{1, 3}} < 0) + \sin(3X_{\cH_{1, 4}}) + 4\sum_{j = 5}^{p_1} \log\{1+(X_{\cH_{1, j}}+1)^2\}/2 + \varepsilon\), where \(\varepsilon\sim\mathcal{P}\), \(X = \Sigma^{-1/2}(Q_{1}, \ldots, Q_{p})^\T\) with \(\Sigma = \diag\{(\rho^{\1(i\neq j)})_{5\times 5}, \ldots, (\rho^{\1(i\neq j)})_{5\times 5}\}\) and \(Q_j\sim \mathcal{P}\) independently for \(j\in[p]\), and \(\cH_{1, j}\) is the \(j\)th element in \(\cH_1\).\label{S3}
\end{enumerate}
The alternative set \(\cH_1\) with size \(p_1\)
is randomly chosen from \([p]\). 
The performance of the proposed MAS is evaluated along with some benchmarks through the comparisons of FDR, averaged power, and running time per replication, where the averaged power is the proportion of signal variables that are correctly identified. All results are obtained with 400 replications with the target FDR level $\alpha=0.2$. 
If not stated otherwise, we consider the predictiveness $V$ as the MSPE in~\ref{M1} and the number of folds of cross-fitting as $K=5$ for the MAS procedure.

\subsection{Choices of agent-X variable}\label{sec:simu:agentX}

First, we investigate the impact of different choices of the agent variable \(\tilde{X}_j\) on the performance of MAS. As outlined in Section~\ref{sec:theory:vim}, we consider the following constructions: (i) setting \(\tilde{X}_j\) as a constant \(c\), where \(c\in\{1, z_{0.95}, z_{0.99}\}\) (denoted MAS-\(c\)), and (ii) setting \(\tilde{X}_j\) as the permuted observations of \(X_j\) (denoted MAS-permu). We use the Model-X knockoff method~\cite{Candes2018} with the approximate second-order Gaussian knockoffs and the lasso coefficient difference statistic as the benchmark (abbreviated as Knockoff).

Table~\ref{tab:00} shows the comparison results under model \ref{S1} with \(p = 300\), \(\delta = 2\), signal proportion \(p_1/p \in \{0.05, 0.1, 0.2\}\), and sample size \(n\) ranging from \(400\) to \(1000\). Here, we apply lasso as the learning algorithm \(\sL\) for MAS.
The results on MAS-\(c\) present a trade-off between FDR control and power---as the constant \(c\) increases from \(1\) to \(z_{0.99}\), the power of MAS improves, but controlling the FDR becomes more challenging, although inflation diminishes with larger sample sizes \(n\). 
In addition, MAS-permu performs similarly to MAS-\(1\). 
These observations are consistent with the discussions in Section~\ref{sec:theory:vim}.  
Overall, setting \(\tilde{X}_j = z_{0.95}\) strikes a favorable balance between power and FDR control under these settings; therefore, we adopt this configuration for the remaining simulations (unless specified otherwise) and recommend it for practice with standardized covariates.
MAS-\(z_{0.95}\) performs better in the sparse signal scenarios, because the learning algorithm lasso itself yields smaller prediction errors when the signals are sparser. Moreover, MAS-\(z_{0.95}\) also outperforms Knockoff in these sparse signal scenarios.

\begin{table}[htbp]\footnotesize\tabcolsep 8pt
  \begin{center}
  \caption{Comparison results of different constructions of agent variables in MAS, with Knockoff as the benchmark, under FDR level \(\alpha = 20\%\) in model \ref{S1}}\label{tab:00}\vspace{-6mm}  
  \end{center}
  \begin{center}
    \resizebox{\linewidth}{!}{
    \begin{tabular}{l *{15}{c}}
    \toprule
          && \multicolumn{2}{c}{MAS-permu} && \multicolumn{2}{c}{MAS-1} && \multicolumn{2}{c}{MAS-\(z_{0.95}\)} && \multicolumn{2}{c}{MAS-\(z_{0.99}\)} && \multicolumn{2}{c}{Knockoff}\\
         \(n\) && FDR & Power && FDR & Power && FDR & Power && FDR & Power && FDR & Power\\
         \hline
         & \multicolumn{15}{l}{\(p_1/p = 0.05\)}\\
         400  &&   17.9 &  61.0 &&  18.5 &  63.1 &&  20.3 &  \ph94.0 &&  20.3 &  \ph99.4 &&   \ph2.5 &  26.9\\
         500  &&   19.2 &  78.9 &&  18.3 &  79.3 &&  18.6 &  \ph98.4 &&  18.7 & 100.0 &&   \ph9.9 &  70.2\\
         700  &&   18.1 &  93.0 &&  18.7 &  93.9 &&  18.3 &  \ph99.8 &&  19.2 & 100.0 &&  15.4 &  93.9\\
         1000  &&   17.7 &  98.6 &&  18.5 &  99.2 &&  19.5 & 100.0 &&  18.4 & 100.0 &&  17.8 &  99.0\\
         & \multicolumn{15}{l}{\(p_1/p = 0.1\)}\\
         400  &&   12.0 &  11.5 &&  13.5 &  11.8 &&  19.6 &  \ph47.9 &&  21.9 &  \ph78.5 &&   \ph0.7 &  18.2\\
         500  &&   16.5 &  21.6 &&  17.1 &  23.5 &&  19.5 &  \ph65.4 &&  20.6 &  \ph90.3 &&   \ph4.9 &  61.4\\
         700  &&   17.7 &  43.2 &&  18.2 &  43.3 &&  19.8 &  \ph85.1 &&  20.3 &  \ph97.9 &&  14.8 &  93.0\\
         1000  &&   19.0 &  63.6 &&  19.3 &  65.3 &&  19.4 &  \ph95.6 &&  20.6 &  \ph99.8 &&  18.1 &  98.6 \\
         & \multicolumn{15}{l}{\(p_1/p = 0.2\)}\\
         400  &&    8.7 &   \ph1.4 &&  10.1 &   \ph1.8 &&  14.2 &   \ph\ph8.5 &&  22.5 &  \ph30.6 &&   \ph0.3 &   \ph6.5\\
         500  &&   11.9 &   \ph4.1 &&  11.8 &   \ph3.7 &&  17.7 &  \ph20.8 &&  22.2 &  \ph49.8 &&   \ph1.6 &  34.1\\
         700  &&   15.3 &   \ph8.5 &&  15.5 &   \ph9.2 &&  19.8 &  \ph39.5 &&  21.7 &  \ph72.3 &&  12.6 &  86.0\\
         1000  &&   16.2 &  16.6 &&  17.7 &  18.2 &&  20.0 &  \ph59.3 &&  21.0 &  \ph87.8 &&  18.1 &  96.4 \\
        \bottomrule
    \end{tabular}}
  \end{center}
\end{table}

\subsection{Ablation experiments}

Then, we elucidate the role of each part in the proposed MAS by comparing it with three benchmarks. The first approach refers to MAS without enhancement of power by specifying $\hat{\beta}_j=1$ in Algorithm~\ref{alg:mas} (denoted by MAS-AgentX). The second one, named as MAS-LOCO, sets $\hat{\beta}_j=1$ and replaces $\hat{\psi}_{j,k}$ with LOCO estimators \(\hat{\varphi}_{j, k} = \pred(\hat{f}_{k, -j}, P_{Z, \cD_{k, 1}^j}) - \pred(\hat{f}_k, P_{Z, \cD_{k, 0}^j})\) in Algorithm~\ref{alg:mas}, where \(\hat{f}_{k, -j}\) is the fitted model using the learning algorithm \(\sL\) based on \(\cD\setminus\cD_k\) with the removal of \(X_j\). The third one is the off-the-shelf BH procedure using p-values derived for the LOCO measures \(\varphi_j\) and denoted by BH-LOCO.
The Model-X knockoff method with the approximate second-order Gaussian knockoffs and the lasso coefficient difference statistic is also adopted as a benchmark (abbreviated as Knockoff).

Table~\ref{tab:02} reports the comparison results under model~\ref{S1} when \(n = 400\), \(\delta = 2\), \(p_1 = \lceil 0.1p\rceil\), and the dimension \(p\) varies from 50 to 250. Here, we apply lasso as the learning algorithm \(\sL\). All methods can control FDR across all the scenarios. The MAS further outperforms MAS-AgentX in terms of higher power due to the efficient data aggregation. 
It is worth noting that the largest enhancement (when $p=250$) is nearly 16\%. 
In comparison with MAS-LOCO and BH-LOCO, MAS-AgentX shows a significant advantage of Agent-X importance measure, which leads to higher empirical power and much faster computation. 
Although BH-LOCO yields a high power compared to MAS-LOCO, we found that there exist strong dependences between p-values used by BH-LOCO. 
This results in BH-LOCO having either no discovery (both FDP and power per replication are zero) or overwhelming discoveries (both FDP and power per replication are close to one). However, the resplit-per-covariate strategy ensures that this problem does not exist for the MAS-based framework. 
More details on the comparison of MAS-LOCO and BH-LOCO can be found in Figure~\ref{fig:supp-tab02} 
of the Appendix~\ref{supp:simul}.
In addition, MAS and MAS-AgentX have comparable computation time to Knockoff, verifying the analysis of computational complexity of using agent-X importance measures mentioned in Section~\ref{sec:algorithm}.

\begin{table}[htbp]\footnotesize\tabcolsep 8pt
  \begin{center}
  \caption{Results of ablation experiments by comparing MAS, MAS-AgentX, MAS-LOCO, and BH-LOCO, together with Knockoff, under FDR level \(\alpha = 20\%\) in model \ref{S1}}\label{tab:02}\vspace{-6mm}
  \end{center}
  \begin{center}
    \resizebox{\linewidth}{!}{
	\begin{tabular}{l *{13}{c}}
    \toprule
          & & \multicolumn{3}{c}{\(p = 50\)} && \multicolumn{3}{c}{\(p = 150\)} && \multicolumn{3}{c}{\(p = 250\)}\\
         Method & & FDR & Power & Time & & FDR & Power & Time & & FDR & Power & Time\\
        \hline
        MAS  &&  17.0 & 100.0 &   \ph0.7 &&  19.2 &  97.7 &   \ph1.9 &&  19.3 &  65.5 &   \ph\ph4.2\\ 
        MAS-AgentX  &&  16.9 & 100.0 &   \ph0.4 &&  20.3 &  94.2 &  \ph1.2 &&  19.8 &  49.1 &   \ph\ph3.1\\ 
        MAS-LOCO  &&  15.5 &  \ph37.9 &  10.0 &&   \ph8.3 &   \ph4.4 &  76.2 &&   \ph6.9 &   \ph1.2 & 478.4\\ 
        BH-LOCO  &&  12.6 &  \ph46.7 &  10.0 &&  13.5 &  15.0 &  76.2 &&  14.6 &  16.2 & 477.6\\
        Knockoff  &&  15.4 &  \ph95.2 &   \ph0.2 &&  15.1 &  91.6 &   \ph1.4 &&   \ph4.4 &  55.1 &   \ph\ph2.7\\ 
         \bottomrule
    \end{tabular}}
  \end{center}
\end{table}

\subsection{Comparisons with competing methods}

\begin{figure}[htbp]
\centering
\includegraphics[scale=0.75]{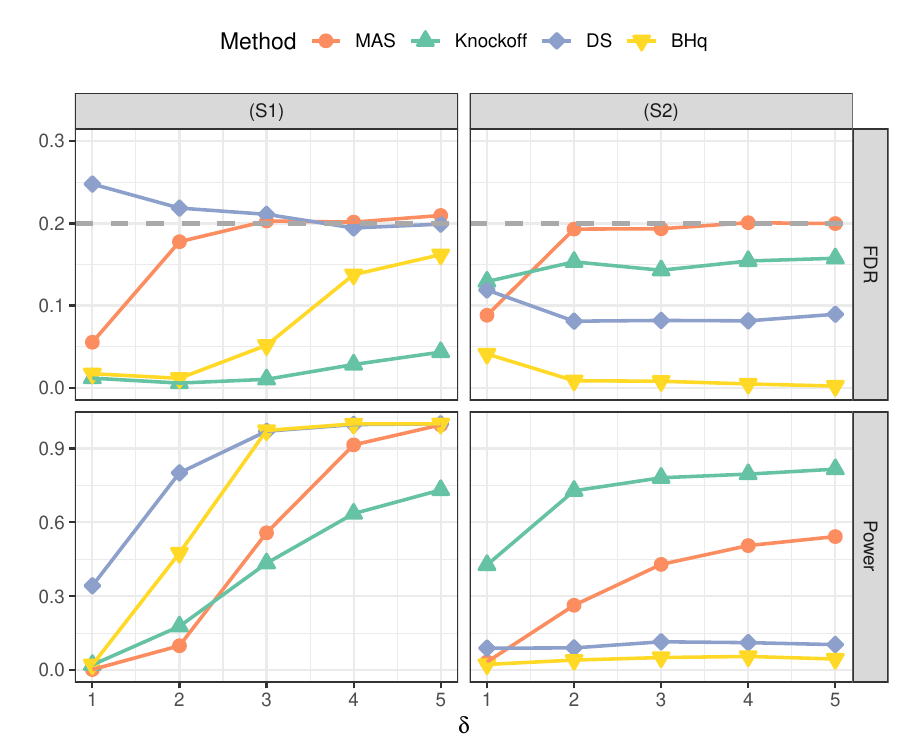}
\caption{Comparison results of MAS, Knockoff~\cite{Candes2018}, DS~\cite{Dai2022ds}, and debiased-lasso-based BHq~\cite{Javanmard2019fdr} under FDR level \(\alpha = 0.2\) (darkgrey, dashed) in models~\ref{S1} and \ref{S2}}
\label{fig:01}
\end{figure}

We next compare MAS with many popular competing methods. For the linear model \ref{S1}, three commonly-used benchmarks are employed, including BH procedure with p-values through debiased lasso \cite{Javanmard2019fdr}, data splitting method \cite{Dai2022ds}, and Model-X knockoff method with the approximate second-order Gaussian knockoffs and the lasso coefficient difference statistic \cite{Candes2018}. 
We refer to these methods above as BHq, DS, and Knockoff, respectively. We specify the learning algorithm \(\sL\) as lasso for MAS. 
The results are displayed in the
left panel of Figure~\ref{fig:01} when \((n, p, p_1) = (600, 400, 80)\) and the magnitude of signals \(\delta\) varies from one to five.  
Both DS and BHq yield FDR control and higher power in comparison to our MAS and Knockoff. This can be expected since both DS and BHq are well designed for linear models, and in the meantime, both model-agnostic methods, MAS and Knockoff, would be endowed with their flexibility. The FDR levels of Knockoff are overly conservative, and in turn its power would be considerably compromised. In contrast, the proposed MAS leads to FDR close to the target level and high power, especially as $\delta$ increases.

We further apply those methods to the non-linear model \ref{S2}. Both DS and BHq are implemented as in the linear model, and the Model-X knockoff method adopts the approximate second-order Gaussian knockoffs and the difference statistic based on random forests. Meanwhile, MAS employs the XGBoost as the learning algorithm \(\sL\). The right panel of Figure~\ref{fig:01} shows the results under \ref{S2} when \((n, p, p_1) = (600, 100, 20)\) and the magnitude of signals \(\delta\) varies from one to five. 
It is evident that the MAS and Knockoff are able to deliver a quite accurate FDR control and perform higher power compared to DS and BHq. This can be understood since MAS and Knockoff are model-free and could capture complex structures. 
We can also observe that MAS is inferior to Knockoff in terms of power, because here, the power of MAS mainly depends on the predictiveness of \(f^*\), i.e., the learning algorithm \(\sL\), while Knockoff is affected by the approximation accuracy of knockoff variables. 
In this Gaussian case with \((n, p) = (600, 100)\), the approximate second-order Gaussian knockoffs are good enough, while the XGBoost used in MAS can be further improved by ensemble methods, as illustrated below in Table~\ref{tab:01}.

We focus on a comparison of MAS and Knockoff in a more complex non-linear model \ref{S3} due to their applicability. 
We keep the setup of Knockoff as in model \ref{S2}.
For MAS, we consider two learning algorithms with different predictive abilities as \(\sL\), the XGBoost and the stacking ensemble framework \cite{WOLPERT1992stacking} integrating the lasso, random forests and XGBoost, which are denoted by MAS-XGBoost and MAS-Ensemble, respectively. 
Table~\ref{tab:01} presents the results under model \ref{S3} with different distributions when \((n, p, p_1) = (600, 100, 20)\) and the dependence parameter \(\rho\) varies from \(0\) to \(0.6\). As illustrated, MAS-XGBoost and MAS-Ensemble perform reasonably well for both settings in terms of FDR control. Meanwhile, MAS-Ensemble delivers power higher than that of MAS-XGBoost (and comparable to Knockoff), which implies that the more predictive capability the learning algorithm in MAS has, the higher detection power MAS can achieve. In contrast, the FDRs of Knockoff are affected by error distributions, and it yields inflated FDRs for a heavy-tailed \(t\) distribution and large $\rho$.

\begin{table}[htbp]\footnotesize\tabcolsep 8pt
  \begin{center}
  \caption{Comparisons of MAS-XGBoost, MAS-Ensemble, and Knockoff under FDR level \(\alpha = 20\%\) in model \ref{S3}}\label{tab:01}\vspace{-2mm}  
  \end{center}
  \begin{center}
    \begin{tabular}{l *{9}{c}}
      \toprule
       && \multicolumn{2}{c}{MAS-XGBoost} && \multicolumn{2}{c}{MAS-Ensemble} && \multicolumn{2}{c}{Knockoff}\\
       \(\rho\) && FDR & Power && FDR & Power && FDR & Power\\ 
       \hline
       && \multicolumn{8}{l}{\(\mathcal{P} \sim \mathcal{N}(0, 1)\)}\\
       0 && 19.8 & 76.3 && 18.3 & 81.0 &&  7.8 & 80.3\\ 
       0.2 && 19.8 & 76.6 && 17.6 & 81.1 &&  8.6 & 80.4\\ 
       0.4 && 18.6 & 75.0 && 18.6 & 81.1 && 10.9 & 78.8\\ 
       0.6 && 19.7 & 71.0 && 19.8 & 81.3 && 13.1 & 67.9\\ 
		  && \multicolumn{8}{l}{\(\mathcal{P} \sim t(3)/\sqrt{3}\)}\\
		  0 && 19.3 & 60.5 && 18.5 & 79.3 &&  8.2 & 80.4\\
		  0.2 && 19.2 & 60.2 && 18.7 & 79.5 && 14.0 & 81.5\\
		  0.4 && 19.3 & 56.6 && 17.5 & 79.3 && 23.4 & 82.0\\
		  0.6 && 17.8 & 46.4 && 20.4 & 77.6 && 29.5 & 76.8\\
      \bottomrule
    \end{tabular}
  \end{center}
\end{table}

In summary, the agent-X importance measures are more suitable for the multiple testing problem than the LOCO measures, and the choice of \(\tilde{X}_j = z_{0.95}\) yields valid FDR control under general settings while providing reasonably high detection power. Furthermore, after determining \(\tilde{X}_j\), the power of MAS mainly depends on the predictive capability of the learning algorithm \(\sL\) (i.e., the predictiveness of \(f^*\)). A learning algorithm \(\sL\) that has more predictive capability brings higher detection power for MAS, 
so ensemble methods are recommended to be used.
On the other hand, compared with the Model-X knockoff framework, which needs to model the joint distribution of covariates accurately, MAS is more recommended in high-dimensional sparse problems, where leveraging structural assumptions for regression is often easier than modeling the joint distribution.

\subsection{Structure-powered enhancements of MAS}

At last, we provide some numerical results on structure-powered enhancement in Section~\ref{sec:ext}.
The data are generated from model \ref{S1}, where the first \(p_1 = \lceil 0.2p\rceil\) variables are taken as \(\cH_1\) to simulate the ordered structure. To compare the performance of the covariate-adaptive importance, we employ random forests as the learning algorithm \(\sL\) for all MAS proecedures below. Then, we consider the MAS with the prior information (denoted MAS-PI), i.e., constructing \(\beta_j = \{\pred(f^*, P_{Z^{\cQ_j}}) + \pred(f^*, P_{Z^j})\}/2\), where $\pred(f^*, P_{Z^{\cQ_j}})$ is the ordered predictiveness measure as introduced in Section~\ref{sec:ext:prior}, and the MAS with covariate-adaptive importance (denoted MAS-CA). 
We also implement Algorithm \ref{alg:mas} with \(\beta_j = \pred(f^*, P_{Z^j})\) and \(\beta_j=1\) as benchmarks, referred to as MAS and MAS-AgentX, respectively.
Table~\ref{tab:03} lists the corresponding performance under the configuration \((n, \delta) = (600, 4)\) and \(p\in\{100, 150, 200\}\).
We can see that the power improvement of the structure-powered methods (MAS-PI and MAS-CA) over MAS is significant.

\begin{table}[htbp]\footnotesize\tabcolsep 8pt
  \begin{center}
  \caption{Comparisons of various power enhancements in MAS framework under model \ref{S1}}\label{tab:03}\vspace{-2mm}  
  \end{center}
  \begin{center}
    \begin{tabular}{c *{12}{c}}
        \toprule
         && \multicolumn{2}{c}{MAS-AgentX} && \multicolumn{2}{c}{MAS} && \multicolumn{2}{c}{MAS-PI} && \multicolumn{2}{c}{MAS-CA} \\
        \(p\) && FDR & Power && FDR & Power && FDR & Power && FDR & Power\\
        \hline
        100  &&   14.2 &  12.4 &&  16.4 &  22.6  &&   9.4 &  43.4  &&  15.2 &  70.9 \\ 
        150  &&    \ph5.8 &   \ph1.4  &&   \ph7.9 &   \ph2.7 &&   \ph3.3 &  12.6 &&   \ph6.1 &  33.4\\
        200  &&    \ph5.4 &   \ph0.6 &&   \ph4.3 &   \ph0.6  &&   \ph1.3 &   \ph5.1  &&   \ph1.4 &  11.6\\
        \bottomrule
    \end{tabular}
  \end{center}
\end{table}

We also implement experiments of selection in the classification task, which are relegated to the Appendix~\ref{supp:simul}. 

\subsection{Real data analysis}

We apply the proposed MAS to detect mutations in the Human Immunodeficiency Virus Type 1 (HIV-1) that are associated with drug resistance. The dataset was analyzed in \cite{HIV2006,Barber2015knockoff,Dai2022ds}, consisting of drug resistance measurements and genotype information from samples of HIV-1. We focus on the data set for resistance to protease inhibitors (PIs).

The response \(Y\) records the log-fold-increase of lab-tested drug resistance, while the covariate \(X_{j} \in\{0, 1\}\) indicates the presence or absence of the \(j\)th mutation. Each drug is analyzed separately. 
The data is preprocessed by first removing the samples with missing drug resistance measurements and then excluding the mutations that appear less than three times in the samples for that drug.
We consider a linear model with no interactions and treat the existing treatment-selected mutation (TSM) panels as well-approximated ground truth. 

We compare the MAS method with three competitors, including the Model-X knockoff method (Knockoff)~\cite{Candes2018}, the multiple data splitting (with 50 splits, DS) in \cite{Dai2022ds}, and the BH method with p-values derived from least squares (BHq). We set the learning algorithm of MAS as the lasso. Except BHq, other methods involve randomness due to data splitting in MAS, DS or knockoff generation in Knockoff. 
Hence, we rerun those mentioned methods for 100 times to make comparisons objectively. 
The nominal FDR level is taken as \(\alpha = 0.2\). 

\begin{figure}[htbp]
  \centering
  \includegraphics[scale=0.75]{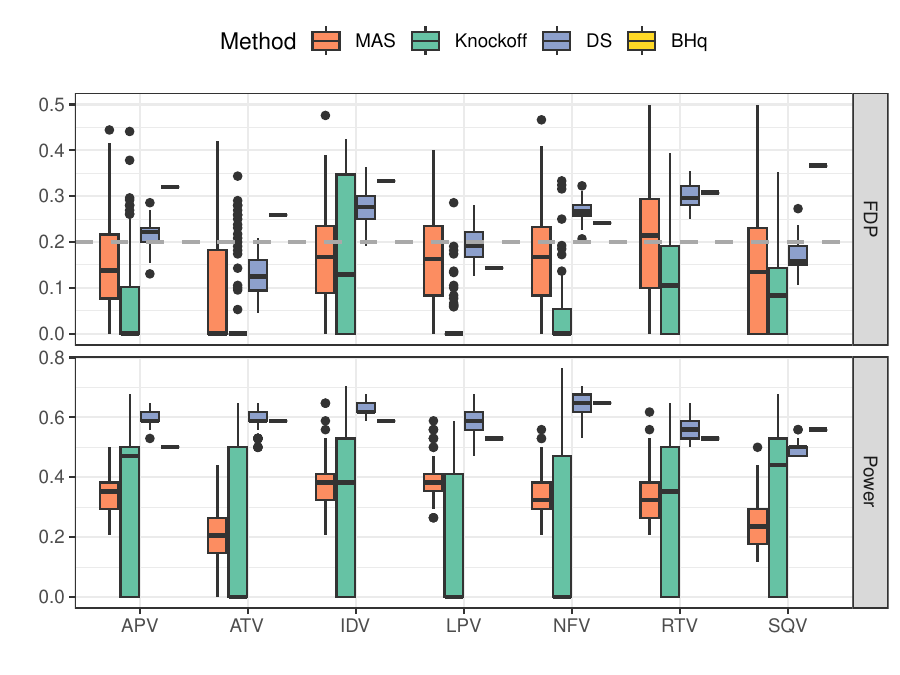}
  \caption{Box plots of FDP and Power of MAS, Knockoff, DS and BHq for detecting mutations associated with seven drug resistance. The nominal FDR level is \(\alpha = 0.2\) (grey dashed line)}
\label{fig:real}
\end{figure}

Figure~\ref{fig:real} depicts the box plots of false discovery proportion (FDP) and proportion of detected signals (Power) over all times for seven drugs. 
We observe that MAS and Knockoff outperform DS and BHq in terms of FDR control across all scenarios. The DS results in inflated FDR for drugs APV, IDV, NFV and RTV. Meanwhile, MAS leads to comparable detection power to Knockoff, but MAS's power is more stable over all drugs. It implies that MAS tends to produce more reliable and replicable findings. 

\section{Concluding Remarks}

This paper introduces a general framework, MAS, to achieve FDR control in a model-agnostic perspective, which is particularly attractive as it requires no prior knowledge on models or distributions, and is conceptually simple and easy to implement. 
We demonstrate both theoretically and empirically that the MAS allows to asymptotically control the FDR in different models with different machine or statistical learning algorithms.

Several directions for further developments are worthwhile to consider. It is noted that MAS proposes to directly apply the predictiveness or integrate prior structures to construct the enhancement parameter $\beta_j$ to improve the SNR. It is worth further study whether there is a more general way to build $\beta_j$ by maximizing power, which corresponds to the goal of identifying as many important variables as possible. The second concern is that we design the covariate-adaptive importance under some specific algorithms in Section \ref{sec:ext:covariate}. This importance demonstrates advantages numerically in terms of high power, but it warrants further research on rigorous theory. 
Finally, we illustrate in Remark~\ref{rem:correlated-covariates} that agent-X importance can retain the signal for strong correlated variables. Further work could investigate the extent to which this behavior persists under nonlinear regression functions and other predictiveness measures, together with its implications for estimation and variable selection under strong covariate dependence. Exploring connections with decorrelated variable importance~\cite{Verdinelli2024decorrelated} may also help characterize these properties.

% Moreover, we focus on the variable importance measures in the form of LOCO and its variants in the MAS framework. 
% Accommodating other developments of variable importance, such as the decorrelated importance~\cite{Verdinelli2024decorrelated}, which is motivated by the powerless performance of LOCO when strong correlations exist between variables, is also worth future exploration.

% \section*{Acknowledgements}
% This work was supported by National Natural Science Foundation of China (Grant No. xxxxxxxx).

\clearpage
\appendix

\section{Agent-X Importance Measure}\label{supp:agentX}
  
  \subsection{Asymptotic linearity}
  
  The proof of Proposition~\ref{prop:asy-linear} 
  is a direct counterpart of Theorem~2 in \cite{Williamson2023vim}. We revisit some notations and conditions. Let \(\cM\) be a rich class of distributions including \(\{P_Z, P_{Z^1}, \ldots, P_{Z^p}\}\), and \(\cF\) be a rich class of functions from \(\cX\) to \(\cY\) equipped with a norm \(\|\cdot\|_\cF\). 
  We define the linear space \(\cH = \{c(P_1 - P_2): c\in[0, \infty), P_1, P_2\in\cM\}\) generated by \(\cM\). And for \(h\in\cH\), say \(h = c(P_1 - P_2)\), we define the supremum norm as \(\|h\|_\infty = c\,\sup_z|F_1(z) - F_2(z)|\), where \(F_1\) and \(F_2\) are the cumulative distribution functions of \(P_1\) and \(P_2\), respectively. 
  Furthermore, we define the random function \(\hat{g}_{k, P}: z \mapsto \dot{\pred}(\hat{f}_k, P; \delta_z - P) - \dot{\pred}(f^*, P; \delta_{z} - P)\). Finally, we consider the following conditions:
  
  \begin{enumerate}[label={(C\arabic*)}]
    \itemindent10pt
      \item (Rate of convergence)\(\|\hat{f}_k - f^*\|_\cF = o_p(n^{-1/4})\) for each \(k\in[K]\).\label{A3}
      \item (Weak consistency)\(\int \{\hat{g}_{k, P}(z)\}^2\d P(z) = o_p(1)\) for any \(P\in\{P_Z, P_{Z^1}, \ldots, P_{Z^p}\}\) and each \(k\in[K]\).\label{A4}
     \item \label{A1} (Optimality) There exists some constant \(C_f > 0\) such that, for each sequence \(f_1, f_2, \ldots \in \cF\) satisfying \(\|f_\ell - f^*\|_\cF \rightarrow 0\), \(|\pred(f_\ell, P_Z) - \pred(f^*, P_Z)| \leq C_f \|f_\ell - f^*\|_\cF^2\) and \(|\pred(f_\ell, P_{Z^j}) - \pred(f^*, P_{Z^j})|\linebreak \leq C_f \|f_\ell - f^*\|_\cF^2\) for each \(\ell\) large enough and \(j\in[p]\). 
      \item \label{A2} (Differentiability) There exists some constant \(\delta > 0\) such that for each sequence \(\epsilon_1, \epsilon_2, \ldots \in\bR\) and \(h, h_1, h_2, \ldots \in \cH\) satisfying that \(\epsilon_j \rightarrow 0\) and \(\|h_j - h\|_\infty \rightarrow 0\), it holds that for any \(P \linebreak\in\{P_Z, P_{Z^1}, \ldots, P_{Z^p}\}\), 
      \[
          \sup_{f\in\cF: \|f - f^*\|_\cF < \delta} \left|\frac{\pred(f, P + \epsilon_j h_j) - \pred(f, P)}{\epsilon_j} - \dot{\pred}(f, P; h_j)\right| \rightarrow 0.
      \]
  \end{enumerate}
  
  Condition~\ref{A3}--\ref{A4} corresponds to Assumption B1--B2 of \cite{Williamson2023vim}, and are conditions on the convergence of \(\hat{f}_k\) to \(f^*\). Conditions~\ref{A1}--\ref{A2} are similar to Assumptions A1--A2 of \cite{Williamson2023vim}.
  Condition~\ref{A1} ensures that there is no first-order contribution that results from estimation of \(f^*\). 
  In Example~\ref{M1}, 
  we have \(|\pred(f, P_Z) - \pred(f^*, P_{Z})| = \E\{f(X) - f^*(X)\}^2\), and thus Condition~\ref{A1} holds with \(C_f = 1\). Other examples can be discussed similarly.
  Condition~\ref{A2} is a form of locally uniform Hadamard differentiability of \(P\mapsto \pred(f^*, P)\) at \(P\in\{P_Z, P_{Z^1} \ldots, P_{Z^p}\}\) in a neighborhood of \(f^*\).
  Due to the linearity of \(\pred(f, P)\) on \(P\) in Example~\ref{M1}--\ref{M3}, 
  Condition~\ref{A2} holds for any \(\delta > 0\). 
  See more detailed discussions in~\cite{Williamson2023vim} for similar assumptions. 
  
  \begin{proof}[Proof of Proposition~\ref{prop:asy-linear}]
          The nature of \(\hat{\psi}_j\) for estimating \(\psi_j = \pred(f^*, P_{Z^j}) - \pred(f^*, P_Z)\) is doing estimation of \(\pred(f^*, P_{Z^j})\) and \(\pred(f^*, P_Z)\) with independent data.
      Following the process of \cite{Williamson2023vim}, we can write 
      \begin{equation*}\label{eq:pm-asymp}
          \begin{split}
              &K^{-1}\sum_{k\in[K]} \pred(\hat{f}_k, P_{Z^j, \cD_{k, 1}^j}) - \pred(f^*, P_{Z^j}) = \dot{\pred}(f^*, P_{Z^j}; P_{Z^j, \cD_1^j} - P_{Z^j}) + (2/n)^{1/2} R_{n, j, 1},\\
              &K^{-1}\sum_{k\in[K]} \pred(\hat{f}_k, P_{Z, \cD_{k, 0}^j}) - \pred(f^*, P_Z) = \dot{\pred}(f^*, P_Z; P_{Z, \cD_{0}^j} - P_Z) + (2/n)^{1/2}R_{n, j, 0},\\
          \end{split}
      \end{equation*} 
      where \(R_{n, j, 1} = o_p(1)\) and \(R_{n, j, 0} = o_p(1)\) are the residuals.
      Then, 
      \begin{equation}\label{eq:psi-asymp-theory}
          \hat{\psi}_j - \psi_j = \frac{2}{n}\sum_{Z_i\in\cD_1^j} \nu_j(Z_i) - \frac{2}{n}\sum_{Z_i\in \cD_0^j} \nu_0(Z_i) + (2/n)^{1/2}(R_{n, j, 1} - R_{n, j, 0}).
      \end{equation}
      The conclusion follows.
  \end{proof}
    
  To satisfy the requirement of convergence rate of \(R_{n, j}\) in Assumption~\ref{asmp:vim}, we can consider the following assumptions.
  \begin{enumerate}[label={(C\arabic*')}]
      \itemindent15pt
      \item \(\|\hat{f}_k - f^*\|_\cF = o_p\{(n\log p)^{-1/4}\}\) for each \(k\in[K]\).\label{A3-1}
      \item \(\int \{\hat{g}_k(z)\}^2 \d P(z) = o_p\{(\log p)^{-1}\}\) for any \(P\in\{P_Z, P_{Z^1}, \ldots, P_{Z^p}\}\) and each \(k\in[K]\).\label{A4-1}
  \end{enumerate}
  
  \begin{proposition}\label{prop:psi-AN}
      Suppose Assumption~\ref{A3-1}--\ref{A4-1} and \ref{A1} hold for \(\pred\) chosen from \ref{M1}--\ref{M3}. 
      Then, Assumption~\ref{asmp:vim}(b) 
      holds.
  \end{proposition}
  
  Proposition~\ref{prop:psi-AN} is a corollary of Proposition~\ref{prop:asy-linear}. 
  We omitted its proof here. 

    \subsection{Non-negativeness of agent-X importance measures}
  
  In this section, we provide the proof of Proposition~\ref{prop:psi0} and the verification of conditions in Proposition~\ref{prop:psi0} for \ref{M1}--\ref{M3}.

  \begin{proof}[Proof of Proposition~\ref{prop:psi0}]
    By the linearity of \(\pred\) on \(P\), we have \(\dot{\pred}(f, P_Z; h) = \pred(f, h)\) for any \(h\in\cH\).
    Then,  
    \[
      \psi_j = \pred(f^*, P_{Z^j}) - \pred(f^*, P_Z) =  \dot{\pred}(f^*, P_Z; P_{Z^j} - P_Z).
    \]
    According to the assumption \(P_{Z^j} \in \cP_{0, j}\), we have \(\psi_j = 0\) for \(P_Z \in\cP_{0, j}\) (i.e., \(Y\indep X_j \mid X_{-j}\)) by the condition (a), and \(\psi_j \geq 0\) otherwise by the condition (b).    
  \end{proof}
  
  \begin{proof}[Verification of conditions in Proposition~\ref{prop:psi0} for \ref{M1}--\ref{M3}]
    For \ref{M1}--\ref{M3}, the predictiveness function \(\pred\) is a linear functional of \(P\), so we have \(\dot{\pred}(f^*, P_Z; P_0 - P_Z) = \pred(f^*, P_0 - P_Z) = \pred(f^*, P_0) - \pred(f^*, P_Z)\). 
  Note that \(f^*(x) = \E(Y\mid X = x)\) in \ref{M1}--\ref{M3}.
  Letting \(W\) be some random variable such that \(P_0 = P_{\{Y, (X_{-j}, W)\}}\), we next verify conditions (a)--(b) for each \(\pred\) respectively.
  
  For \(\pred\) specified as \ref{M1}, 
  we have
  \[\begin{split}
      &\dot{\pred}(f^*, P_Z; P_0 - P_Z) = \pred(f^*, P_0) - \pred(f^*, P_Z)\\
      &\quad = \E\{Y - f^*(X_{-j}, W)\}^2 - \E\{Y - f^*(X_{-j}, X_j)\}^2\\
      &\quad = \E\{f^*(X_{-j}, X_j) - f^*(X_{-j}, W)\}^2 + \E[\{f^*(X_{-j}, X_j) - f^*(X_{-j}, W)\}\{Y - f^*(X_{-j}, X_j)\}]\\
      &\quad = \E\{f^*(X_{-j}, X_j) - f^*(X_{-j}, W)\}^2,\\
  \end{split}\]
  where the last equality holds due to \(\E(Y\mid X, W) = \E(Y\mid X)\). Then, when \(P_Z\in\cP_{0, j}\), i.e., \(X_j\indep Y\mid X_{-j}\) (and thus \(f^*(x_{-j}, x_j) = \E(Y\mid X_{-j} = x_{-j})\)), \(\dot{\pred}(f^*, P_Z; P_0 - P_Z) = 0\); when \(P_Z\not\in\cP_{0, j}\), \(\dot{\pred}(f^*, P_Z; P_0 - P_Z) \geq 0\).
  
  For \(\pred\) specified as \ref{M2}, 
  we have
  \[\begin{split}
      &\dot{\pred}(f^*, P_Z; P_0 - P_Z) = \pred(f^*, P_0) - \pred(f^*, P_Z)\\
      &\quad = \E\left(Y\log\{f^*(X_{-j}, {X}_j)/f^*(X_{-j}, W)\} + (1-Y)\log[\{1-f^*(X_{-j}, {X}_{j})\}/\{1-f^*(X_{-j}, W)\}]\right)\\
      &\quad = \E\left(f^*(X_{-j}, X_j) \log\{f^*(X_{-j}, {X}_j)/f^*(X_{-j}, W)\}\right.\\ 
      &\qquad + \left.\{1-f^*(X_{-j}, X_j)\}\log[\{1-f^*(X_{-j}, {X}_{j})\}/\{1-f^*(X_{-j}, W)\}]\right)\\
      &\quad \geq 0,\\
  \end{split}\]
  where 
  the last inequality holds due to \(\log x \leq x-1\) and the equality holds only when \(\P\{f^*(X_{-j}, X_j) = f^*(X_{-j}, W)\} = 1\), implied by \(P_Z\in\cP_{0, j}\).
  
  By similar derivations, we have \(\dot{\pred}(f^*, P_Z; P_0 - P_Z) \geq 0\) for \ref{M3}.
  The justification ends.       
  \end{proof}
  
  \section{Resplit-Per-Covariate Strategy}\label{supp:dep}

\subsection{Generalization of resplit-per-covariate strategy}

In Section~\ref{sec:theory:random}, 
we have introduced the random variable \(\gamma_{j, i} = \1(Z_i\in\cD_1^j)\) to represent the randomness of the resplitting process and rewritten the asymptotic linearity as 
\[
\frac{\sqrt{n}}{2} (\hat{\psi}_j - \psi_j) = \frac{1}{\sqrt{n}}\sum_{i\in[n]}\{\gamma_{j, i}\nu_j(Z_i) - (1-\gamma_{j, i})\nu_0(Z_i)\} + R_{n, j},
\]
where \(\{\gamma_{j, i}: i\in[n]\}\) are weakly dependent as we have the constraint \(\sum_{i\in[n]} \gamma_{j, i} = n/2\). 
To analyze two agent-X importance estimators jointly, we need to find a set of independent random variables \(\{\xi_{j, i}: i\in[n]\}\) to asymptotically replace \(\{\gamma_{j, i}: i\in[n]\}\).

In this section, we consider a generalization of the resplit-per-covariate strategy, where the samples yield by resplitting could be uneven, i.e., \(\sum_{i\in[n]} \gamma_{j, i} = n_1\), where \(n_1\) is a user-specified integer.
Recalling the asymptotic expression in \eqref{eq:psi-asymp-theory} for \(j\in[p]\), we have
\begin{align}
    \hat{\psi}_{j} - \psi_j &= \frac{1}{n_1}\sum_{i\in\cI_{1}^j} \nu_j(Z_i) - \frac{1}{n_0}\sum_{i\in\cI_{0}^j} \nu_0(Z_i) + n_1^{-1/2}R_{n, j, 1} - n_0^{-1/2}R_{n, j, 0} \nonumber\\
    &= \sum_{i\in[n]} \left\{\frac{\gamma_{j, i}}{n_1}\nu_j(Z_i) - \frac{1-\gamma_{j, i}}{n_0}\nu_0(Z_i)\right\} + n_1^{-1/2}R_{n, j, 1} - n_0^{-1/2}R_{n, j, 0}. \label{eq:psi-asymp-supp}    
\end{align}
Similar to Lemma~\ref{lem:removeDep}, 
we can find a term consisting of independent sums to approximate the major term in the asymptotic expression \eqref{eq:psi-asymp-supp}. 

\begin{lemma}\label{supp:lem:removeDep}
    Suppose that Assumption~\ref{asmp:moment} 
    holds. 
    Additionally, we assume \(c_0 \leq  n_1/n \leq 1 - c_0\) for some constant \(c_0 \in (0, 1/2]\).
    Then, there exists \(\{\xi_{j, i}: i \in [n], j \in [p]\}\) satisfying that \(\xi_{j, i}\) across \(i\in[n]\) and \(j\in[p]\) are independent Bernoulli random variables with probability \(n_1/n\), such that 
    \[\begin{split}
        &\P\left(\max_{j\in[p]}|{\bG}_{j} - \tilde{\bG}_{j}| \geq C[(\log p)^{3}/n]^{1/4}\right) \leq  3(\log n)^{-1},
    \end{split}\]
    where \(C > \{4C_{\sigma}/(1-a)\}\sqrt{(1-c_0)/c_0}\) for some \(a\in(0, 1)\), and \(\bG_{j}\) and \(\tilde{\bG}_{j}\) are defined as
    \[\begin{split}
        {\bG}_{j} &= \sqrt{\frac{n_1n_0}{n}}\sum_{i \in[n]} \left\{\frac{\gamma_{j, i}}{n_1}\nu_j(Z_i) - \frac{1-\gamma_{j, i}}{n_0}\nu_0(Z_i)\right\},\\
        \tilde{\bG}_{j} &= \sqrt{\frac{n_1n_0}{n}}\sum_{i \in [n]} \left\{\frac{\xi_{j, i}}{n_1}\nu_j(Z_i) - \frac{1-\xi_{j, i}}{n_0}\nu_0(Z_i)\right\}.
    \end{split}\]  
\end{lemma}

\subsection{Proof of Lemma~\ref{lem:removeDep} 
and \ref{supp:lem:removeDep}}

The resplitting is done for each fold \(k\in[K]\) and each covariate index \(j\in[p]\) independently. Thus, \(\gamma_{j, i_1}\) and \(\gamma_{j, i_2}\) are dependent only when \({i_1}\) and \({i_2}\) are in the same fold. 
We can remove the dependence through Binomialization, a technique similar to Poissonization in~\cite{Vaart1996}, in each fold separately, and merge the independent terms from different folds. 

Without loss of generality, we present the analysis of one fold, where for notational convenience we assume this fold has \(n\) samples, and \(\cD_0^j\) and \(\cD_{1}^j\) with \(n_0 = |\cD_0^j|\) and \(n_1 = |\cD_1^j|\) are the yielded subsets by a resplitting for the index \(j\). Then, the asymptotic expression in \eqref{eq:psi-asymp-supp} holds.

Now, we prove Lemma~\ref{supp:lem:removeDep} (and Lemma~\ref{lem:removeDep} 
as a special case of \(n_1 = n/2\) and \(c_0 = 1/2\)). 
First, we consider constructing the random variables \(\{\xi_{j, i}: i\in[n]\}\) for the index \(j\), and the construction of random variables for other indices is the same. 
Let 
\[\begin{split}
    &\omega_{n, j}^{(1)} \sim \mathrm{Multinomial}(1, (n^{-1}, \ldots, n^{-1}))\quad \text{and}\quad \omega_{n, j}^{(i)}\mid \sum_{k = 1}^{i-1} \omega_{n, j}^{(k)} \sim \mathrm{Multinomial}(1, \mu_{i}),\ i = 2, \ldots, n,
\end{split}\] 
where \(\mu_i\) is an \(n\)-dimensional vector with nonzero probability \((n-i+1)^{-1}\) at the locations where \(\sum_{k = 1}^{i-1}\omega_{n, j}^{(k)}\) are zeros. 
All \(\omega_{n, j}^{(i)}\) are generated independent of \(N_{n, j}\sim\mathrm{Binomial}(n, n_1/n)\).
Then, we set 
\begin{equation}\label{eq:supp:decouple}
    {\gamma}_{j} = (\gamma_{j, 1}, \ldots, \gamma_{j, n})^\T = \sum_{i = 1}^{n_1} \omega_{n, j}^{(i)}, \quad {\xi}_{j} = (\xi_{j, 1}, \ldots, \xi_{j, n})^\T = \sum_{i = 1}^{N_{n, j}} \omega_{n, j}^{(i)}.
\end{equation}
Define \({\bG}_{j}\) using \({\gamma}_{j}\) and \(\tilde{\bG}_{j}\) using \({\xi}_{j}\), respectively, i.e., 
\[\begin{split}
    {\bG}_{j} &= \sqrt{\frac{n_1n_0}{n}}\sum_{i \in[n]} \left\{\frac{\gamma_{j, i}}{n_1}\nu_j(Z_i) - \frac{1-\gamma_{j, i}}{n_0}\nu_0(Z_i)\right\},\\ \tilde{\bG}_{j} &= \sqrt{\frac{n_1n_0}{n}}\sum_{i \in [n]} \left\{\frac{\xi_{j, i}}{n_1}\nu_j(Z_i) - \frac{1-\xi_{j, i}}{n_0}\nu_0(Z_i)\right\}.
\end{split}\]
Since for any \((x_1, \ldots, x_n)\in\{0, 1\}^n\),
\[
    \P(\xi_{j, 1} = x_1, \ldots, \xi_{j, n} = x_n\mid N_{n, j} = k) = \left\{\begin{array}{ll}
        1/\binom{n}{k}, & \text{if } \sum_{i = 1}^n x_i = k,\\
        0, & \text{otherwise},\\
    \end{array}\right.
\]
Lemma~\ref{lem:binomialization} implies \(\xi_{j, 1}, \ldots, \xi_{j, n}\) are independent Bernoulli with probability \(n_1/n\).
Due to the independent construction of \(\{\xi_{j, i}: i\in[n]\}\) across \(j\in[p]\), we get \(\{\xi_{j, i}: i\in[n], j\in[p]\}\) are independent Bernoulli random variables with probability \(n_1/n\).

Next, we characterize the distance between \({\bG}_{j}\) and \(\tilde{\bG}_{j}\).
By definition, 
\[
    {\bG}_{j} - \tilde{\bG}_{j} = (n_1n_0/n)^{1/2}\sum_{i\in[n]} \left(\gamma_{j, i} - \xi_{j, i}\right) \left\{\nu_0(Z_i)/n_0 + \nu_{j}(Z_i)/n_1\right\} := Q_{n, j}.
\]
Note that the absolute difference vector \(|{\gamma}_{j} - {\xi}_{j}|\) is the sum of \(|n_1 - N_{n, j}|\) variable vectors \(\omega_{n, j}^{(i)}\). 
Let \(\cJ_{n, j}\) be the set of indices \(i\in[n]\) such that \(|\gamma_{j, i} - \xi_{j, i}| = 1\).  
We then have \(\gamma_{j, i} - \xi_{j, i} = \mathrm{sign}(n_1 - N_{n, j}) \1(i\in\cJ_{n, j})\). 
Thus, 
\begin{equation}\label{eq:supp:Unj}
    Q_{n, j} = \mathrm{sign}\left(n_1 - N_{n, j}\right) n^{-1/2}\sum_{i\in\cJ_{n, j}} \left\{\nu_0(Z_i)(n_1/n_0)^{1/2} + \nu_{j}(Z_i)(n_0/n_1)^{1/2}\right\}.
\end{equation}
Define \(\zeta_{ij} = \nu_0(Z_i)(n_1/n_0)^{1/2} + \nu_j(Z_i)(n_0/n_1)^{1/2}\) and \(\varsigma_j^2 = \E(\zeta_{ij}^2)\) for \(j\in[p]\), where the variance \(\varsigma_j^2\) satisfies 
\begin{equation}\label{eq:S1}
    \varsigma_j^2 \leq 2(n_1/n_0)\E\{\nu_0(Z)\}^2 + 2(n_0/n_1)\E\{\nu_j(Z)\}^2 \leq \{4(1-c_0)/c_0\}\sigma_j^2.
\end{equation}
Let \(A_n = 2\{(1-c_0)/c_0\}^{1/2}(Mn\log n)^{1/\theta}\) for \(\theta > 4\) and \(\epsilon = C[(\log p)^3/n]^{1/4}\) for \(C>\{4C_{\sigma}/(1-a)\}\sqrt{(1-c_0)/c_0}\) and some \(a \in(0, 1)\). 
We do the decomposition
\[\begin{split}
    \zeta_{ij} &= \zeta_{ij}\1(|\zeta_{ij}|\leq A_n) - \E\{\zeta_{ij}\1(|\zeta_{ij}|\leq A_n)\} + \zeta_{ij}\1(|\zeta_{ij}| > A_n) - \E\{\zeta_{ij}\1(|\zeta_{ij}| > A_n)\}\\
    &:= \zeta_{ij, 1} + \zeta_{ij, 2}.
\end{split}\]
We have
\[\begin{split}
    \P\left(\max_{j\in[p]} |Q_{n, j}| \geq \epsilon\right) &\leq \P\left(\max_{j\in[p]}\left|\frac{1}{\sqrt{n}}\sum_{i\in\cJ_{n, j}}\zeta_{ij}\right| \geq \epsilon\right)\\
    &\leq \P\left(\max_{j\in[p]}\left|\frac{1}{\sqrt{n}}\sum_{i\in\cJ_{n, j}}\zeta_{ij, 1}\right| \geq (1-a)\epsilon\right) + \P\left(\max_{j\in[p]}\left|\frac{1}{\sqrt{n}}\sum_{i\in\cJ_{n, j}}\zeta_{ij, 2}\right| \geq a\epsilon\right)\\
    &=: P_1 + P_2.
\end{split}\]
For \(P_1\), by Lemma~\ref{lem:Bernstein},
\[\begin{split}
    P_1 &\leq 2\sum_{j\in[p]} \P\left(\frac{1}{\sqrt{n}}\sum_{i\in\cJ_{n, j}}\zeta_{ij, 1} \geq (1-a)\epsilon\right)\\
    &\leq 2 \sum_{j\in[p]}\E\left\{\P\left(\frac{1}{\sqrt{n}}\sum_{i\in\cJ_{n, j}}\zeta_{ij, 1} \geq (1-a)\epsilon\mid \cJ_{n, j}\right)\right\}\\
    &\leq 2\sum_{j\in[p]}\E\left\{\exp\left(-\frac{\sqrt{n}\epsilon^2}{\{2\varsigma_j^2/(1-a)^2\}|\cJ_{n, j}|/\sqrt{n} + [2/\{3(1-a)\}]A_n\epsilon}\right)\right\}.\\
\end{split}\]
Let \(c_1 = 2/(1-a)^2\) and \(c_2 = 2/\{3(1-a)\}\).
Note that \(|\cJ_{n, j}| = |N_{n, j} - n_1|\) is the size of \(\cJ_{n, j}\), and \(N_{n, j}\sim \mathrm{Binomial}(n, n_1/n)\). 
By Hoeffding's inequality,  
\begin{equation}\label{eq:J}
    \P\left(|\cJ_{n, j}| \geq t\right) = \P\left(|N_{n, j} - n_1| \geq t\right) \leq 2\exp\left(-{2t^2}/{n}\right).
\end{equation}
So we make a truncation for \(|\cJ_{n, j}|\),
\[\begin{split}
    &\E\left\{\exp\left(-\frac{\sqrt{n}\epsilon^2}{\{2\varsigma_j^2/(1-a)^2\} |\cJ_{n, j}|/\sqrt{n} + [2/\{3(1-a)\}] A_n\epsilon}\right)\right\}\\
    &\quad = \E\left\{\exp\left(-\frac{\sqrt{n}\epsilon^2}{(c_1\varsigma_j^2) |\cJ_{n, j}|/\sqrt{n} + c_2A_n\epsilon}\right)\1\left(|\cJ_{n, j}|\leq \sqrt{n\log p}\right)\right\}\\
    &\hspace*{2cm} + \E\left\{\exp\left(-\frac{\sqrt{n}\epsilon^2}{(c_1\varsigma_j^2) |\cJ_{n, j}|/\sqrt{n} + c_2A_n\epsilon}\right)\1\left(|\cJ_{n, j}|> \sqrt{n\log p}\right)\right\}.\\
    &\quad \leq \exp\left(-\frac{\sqrt{n}\epsilon^2}{c_1\varsigma_j^2\sqrt{\log p} + c_2 A_n\epsilon}\right) + \exp\left(-\frac{\epsilon^2}{c_1\varsigma_j^2 + c_2 A_n\epsilon/\sqrt{n}}\right)\P\left(|\cJ_{n, j}| > \sqrt{n\log p}\right)\\
    &\quad \leq \exp\left(-\frac{\sqrt{n}\epsilon^2}{c_1\varsigma_j^2\sqrt{\log p} + c_2 A_n\epsilon}\right) + 2p^{-2}\exp\left(-\frac{\epsilon^2}{c_1\varsigma_j^2 + c_2 A_n\epsilon/\sqrt{n}}\right)\\
\end{split}\]
where the first inequality holds due to \(|\cJ_{n, j}| \leq n\), and the last inequality holds due to \eqref{eq:J}. 
Then, 
\[
    P_1 \leq 2p\exp\left(-\frac{\sqrt{n}\epsilon^2}{c_1\varsigma_j^2 \sqrt{\log p} + c_2 A_n\epsilon}\right) + 4p^{-1}\exp\left(-\frac{\epsilon^2}{c_1\varsigma_j^2 + c_2 A_n\epsilon/\sqrt{n}}\right) \leq 6p^{-1},
\]
where the last inequality holds by \(A_n(\log p)^{1/4}/n^{1/4} \rightarrow 0\) as \(n, p\rightarrow\infty\) for \(\theta > 4\).
For \(P_2\), we have 
\[\begin{split}
    \max_{j\in[p]} \sum_{i\in\cJ_{n, j}} |\zeta_{ij, 2}| &\leq \max_{j\in[p]}\sum_{i\in\cJ_{n, j}} |\zeta_{ij}|\1(|\zeta_{ij}| > A_n) + \max_{j\in[p]} |\cJ_{n, j}| \left|\E\left\{\zeta_{ij}\1(|\zeta_{ij}| > A_n)\right\}\right|\\
    &\leq \sum_{i\in[n]} \max_{j\in[p]}|\zeta_{ij}|\1(|\zeta_{ij}| > A_n) + \max_{j\in[p]} |\cJ_{n, j}| \left|\E\left\{\zeta_{ij}\1(|\zeta_{ij}| > A_n)\right\}\right|.
\end{split}\]
By Cauchy--Schwarz inequality and \eqref{eq:S1},
\begin{equation}\label{eq:S2}
    \begin{split}
        |\E\left\{\zeta_{ij}\1(|\zeta_{ij}| > A_n)\right\}| &\leq \{\E\left(\zeta_{ij}^2\right)\P(|\zeta_{ij}| > A_n)\}^{1/2} \leq \varsigma_j \left(\E|\zeta_{ij}|^\theta/A_n^\theta\right)^{1/2}\leq 2cC_{\sigma}(n\log n)^{-1/2},
    \end{split}
\end{equation}
where the last inequality holds by \(\E|\zeta_{ij}|^\theta \leq 2^{\theta - 1}M\{(n_1/n_0)^{\theta/2} + (n_0/n_1)^{\theta/2}\} \leq (2c)^{\theta}  M\) with \(c = \{(1-c_0)/c_0\}^{1/2}\). 
Next, we define an event 
\[
    \cE = \left\{n^{-1/2}\max_{j\in[p]} |\cJ_{n, j}| \left|\E\left\{\zeta_{ij}\1(|\zeta_{ij}| > A_n)\right\}\right| < a\epsilon/2\right\}.
\]
With \(\P(\max_{j\in[p]} |\cJ_{n, j}| \geq t) \leq 2p\exp(-2t^2/n)\) from \eqref{eq:J}, we have 
\begin{equation}\label{eq:S3}
    \E\left\{\max_{j\in[p]} |\cJ_{n, j}|\right\} \leq \sqrt{n/2}\left\{\sqrt{\log (2p)}+\sqrt{\pi}/2\right\}
\end{equation}
by Lemma~\ref{lem:exp-Hoeffding}. Then, by \eqref{eq:S2},
\[\begin{split}
    \P(\cE^c) &\leq \P\left(2cC_{\sigma}(n\log n)^{-1/2} \max_{j\in[p]} |\cJ_{n, j}|/\sqrt{n} \geq a\epsilon/2\right)\\
    &\leq \frac{4cC_{\sigma}(n\log n)^{-1/2}\E\{\max_{j\in[p]}|\cJ_{n, j}|\}}{a\epsilon\sqrt{n}}\\
    &\lesssim \{2\sqrt{2} c C_{\sigma} /a\} n^{-1/4}(\log p)^{-1/4}(\log n)^{-1/2},
\end{split}\]
where the second inequality holds by Markov's inequality, and the third inequality holds due to \eqref{eq:S3}.
Back to \(P_2\),
\[\begin{split}
    P_2 &\leq \P\left(\left\{\max_{j\in[p]} \frac{1}{\sqrt{n}} \sum_{i\in\cJ_{n, j}}|\zeta_{ij, 2}| \geq a\epsilon\right\}\cap \cE\right) + \P\left(\left\{\max_{j\in[p]} \frac{1}{\sqrt{n}} \sum_{i\in\cJ_{n, j}}|\zeta_{ij, 2}| \geq a\epsilon\right\}\cap \cE^c\right)\\
    &\leq \P\left(\frac{1}{\sqrt{n}}\sum_{i\in[n]} \max_{j\in[p]} |\zeta_{ij}|\1(|\zeta_{ij}| > A_n) \geq a\epsilon/2\right) + \P(\cE^c)\\
    &\leq \P\left(\max_{j\in[p]}|\zeta_{ij}| > A_n \text{ for some }i\in[n]\right) +  \{2\sqrt{2} c C_{\sigma} /a\} n^{-1/4}(\log p)^{-1/4}(\log n)^{-1/2}\\
    &\leq n \frac{\E[\max_{j\in[p]} |\zeta_{ij}|^\theta]}{A_n^\theta} + \{2\sqrt{2} c C_{\sigma} /a\} n^{-1/4}(\log p)^{-1/4}(\log n)^{-1/2}\\
    &\leq (\log n)^{-1} + \{2\sqrt{2} c C_{\sigma} /a\} n^{-1/4}(\log p)^{-1/4}(\log n)^{-1/2},
\end{split}\]
where the fourth inequality holds by Markov's inequality, and the last inequality holds by the definition of \(A_n\) and \(\E[\max_{j\in[p]}|\zeta_{ij}|^\theta]\leq (2c)^{\theta} M\).
So, we combine the upper bounds of \(P_1\) and \(P_2\),
\[\begin{split}
    \P\left(\max_{j\in[p]} |Q_{n, j}| \geq \epsilon\right) &\leq 6p^{-1} + (\log n)^{-1} +  \{2\sqrt{2} c C_{\sigma} /a\} n^{-1/4}(\log p)^{-1/4}(\log n)^{-1/2} \leq 3 (\log n)^{-1},
\end{split}\]
where the first inequality holds due to \(A_n(\log p)^{1/4}/n^{1/4} \rightarrow 0\) as \(n, p\rightarrow\infty\) for \(\theta > 4\).
Thus, the conclusion of Lemma~\ref{supp:lem:removeDep} follows.

\section{Proofs}\label{supp:proof}

\subsection{Proof of Theorem~\ref{THM:FDR}}

We consider \(p = O(n^r)\) for some constant \(r > 0\). For \(j\in[p]\), we multiply \(\hat{\psi}_{j}^\B\) in \eqref{eq:psiE} 
by \(\sqrt{n}/2\) as the statistics in theoretical analysis, which would not change the discoveries due to the form of the threshold rule \eqref{eq:thrsh}. 
We define 
\[
    G(t) = p_0^{-1}\sum_{j\in\cH_0}[1 - \Phi\{t/(\sigma_{j}\beta_{j})\}],
\] 
where \(\Phi(t)\) is the cumulative distribution function of a standard Gaussian random variable. 
Let \(t^*\) be a nonrandom sequence related to \(n, p\), such that 
\begin{equation}\label{eq:G}
    G(t^*) = \frac{\alpha p_{1*}}{2(1-\alpha)p_0}.
\end{equation}
Let \(u_{\sigma, 0}, \ell_{\sigma, 0}, u_{\sigma, 1}, \ell_{\sigma, 1}\) be some constants such that $0 < \ell_{\sigma, 0} \leq \sigma_j \leq u_{\sigma, 0} < \infty$ for \(j\in\cH_0\) and \(0 < \ell_{\sigma, 1} \leq \sigma_j \leq u_{\sigma, 1} < \infty\) for \(j\in\cH_1\), according to Assumption~\ref{asmp:moment}(a).
Let \(\ell_{\beta, 0}, \ell_{\beta, 1}, u_{\beta, 0}, u_{\beta, 1}\) be constants such that \(0 < \ell_{\beta, 0} \leq \beta_j \leq u_{\beta, 0} < \infty\) for \(j\in\cH_0\) and \(0 < \ell_{\beta, 1} \leq \beta_j \leq u_{\beta, 1} < \infty\) for \(j\in \cH_1\), according to Assumption~\ref{asmp:beta}(a). 

\begin{lemma}\label{lem:exp}
    Suppose Assumption~\ref{asmp:moment} 
    holds.  Then, 
    \[
        \P\left(\bigcup_{j \in[p]}\left\{\tilde{\bG}_{j} \geq \sigma_j\sqrt{C\log p}\right\}\right) = o(1),
    \]
    where \(\tilde{\bG}_{j} = {n}^{-1/2}\sum_{i\in[n]}\{\xi_{j, i} \nu_j(Z_i) - (1-\xi_{j, i})\nu_0(Z_i)\}\), and \(C > 4/(1 - a)^2\) and \(0 < a < 1\) are constants.
\end{lemma}

\begin{proof}[Proof of Lemma~\ref{lem:exp}]
    Recall that 
    \[
        \tilde{\bG}_{j} = \frac{1}{\sqrt{n}}\sum_{i = 1}^n \left\{\xi_{j, i}\nu_j(Z_i) - (1-\xi_{j, i})\nu_0(Z_i)\right\},
    \]
    where \(\xi_{j, 1}, \ldots, \xi_{j, n}\) are independent Bernoulli random variables with probability \(1/2\). Let \(A_n = (nM\log n)^{1/\theta}\) for \(\theta > 2\). Define 
    \[\begin{split}
        \zeta_{ij} &= \xi_{j, i}\nu_j(Z_i) - (1-\xi_{j, i})\nu_0(Z_i)\\
        &= \zeta_{ij}\1(|\zeta_{ij}|\leq A_n) - \E\{\zeta_{ij}\1(|\zeta_{ij}|\leq A_n)\} + \zeta_{ij}\1(|\zeta_{ij}|> A_n) - \E\{\zeta_{ij}\1(|\zeta_{ij}| > A_n)\}\\
        &:= \zeta_{ij, 1} + \zeta_{ij, 2}.
    \end{split}\]
    For small \(a > 0\), we have 
    \[\begin{split}
        &\P\left(\bigcup_{j\in[p]}\left\{\bG_{N_{n, j}} \geq \sigma_j\sqrt{C\log p}\right\}\right) = \P\left(\bigcup_{j\in[p]}\left\{\frac{1}{\sqrt{n}}\sum_{i = 1}^n (\zeta_{ij, 1} + \zeta_{ij, 2})\geq \sigma_j\sqrt{C\log p}\right\}\right)\\
        &\qquad \leq \P\left(\bigcup_{j\in[p]}\left\{\frac{1}{\sqrt{n}}\sum_{i = 1}^n \zeta_{ij, 1} \geq (1-a)\sigma_j\sqrt{C\log p}\right\}\right)\\
        &\qquad\quad + \P\left(\bigcup_{j\in[p]}\left\{\frac{1}{\sqrt{n}}\sum_{i = 1}^n \zeta_{ij, 2} \geq a \sigma_j\sqrt{C\log p}\right\}\right)\\
        &\qquad := Q_1 + Q_2
    \end{split}\]
    For \(Q_1\), by Lemma~\ref{lem:Bernstein}, 
    \[\begin{split}
        Q_1 &\leq \sum_{j\in[p]}\P\left(\frac{1}{\sqrt{n}}\sum_{i = 1}^n \zeta_{ij, 1} \geq (1-a)\sigma_j\sqrt{C\log p}\right)\\
        &\leq p\exp\left(-\frac{C\log p}{1/(1-a)^{2} + [2\sqrt{C}/\{3\sigma_j(1-a)\}] A_n\sqrt{\log p/n}}\right) = o(1),
    \end{split}\]
    where \(A_n\sqrt{\log p / n} \rightarrow 0\) as \(n\rightarrow \infty\) and \(C > 4/(1-a)^2\). 
    Note that 
    \[\begin{split}
        &\max_{j\in[p]}\sqrt{n}\left|\E\left\{\zeta_{ij}\1(|\zeta_{ij}| > A_n)\right\}\right| \leq \max_{j\in[p]}\sqrt{n\E(\zeta_{ij}^2)\P(\zeta_{ij} > A_n)}\\
        &\qquad \leq (u_{\sigma, 0}\vee u_{\sigma, 1})\sqrt{\frac{n\max_{j\in[p]}\E|\zeta_{ij}|^\theta}{A_n^\theta}} \leq (u_{\sigma, 0}\vee u_{\sigma, 1}) \sqrt{nM} A_n^{-\theta/2} = o\left(\sqrt{\log p}\right),
    \end{split}\]
    where the first inequality holds by Cauchy--Schwarz inequality, and the third inequality holds by Assumption~\ref{asmp:moment}.
    Thus, for large enough \(n, p\), we have
    \[\begin{split}
        Q_2 &\leq \P\left(\bigcup_{j\in[p]} \left\{\sqrt{\frac{2}{n}}\sum_{i = 1}^n |\zeta_{ij}|\1(|\zeta_{ij}| > A_n) \geq (a/2)\sigma_j\sqrt{C\log p}\right\}\right)\\
        &\leq \P\left(\max_{j\in[p]}|\zeta_{ij}| > A_n \text{ for some } i\right)\\
        &\leq \frac{n \E[\max_{j\in[p]}|\zeta_{ij}|^\theta]}{A_n^\theta} \leq (\log n)^{-1} = o(1).
    \end{split}\]
    Thus, we complete the proof.  
\end{proof}

Next, we present the proof of Theorem~\ref{THM:FDR}.

\Step{Step~1}(Residual terms).\quad 
We first show that 
\begin{equation}\label{eq:step1}
    \sqrt{n}(\hat{\psi}_j - \psi_j)\hat{\beta}_j/2 = \tilde{\bG}_{j}\beta_j + o_p\{(\log p)^{-1/2}\}.        
\end{equation}

Combining Lemma~\ref{lem:removeDep} 
and Assumption~\ref{asmp:vim}, 
we have 
\[  
    \frac{\sqrt{n}}{2}(\hat{\psi}_j - \psi_j) = \tilde{\bG}_{j} + Q_{n, j} + R_{n, j}, 
\]
where \(Q_{n, j} = \bG_{j} - \tilde{\bG}_{j}\) satisfies \(\max_{j\in[p]}|Q_{n, j}| = O_p([(\log p)^3/n]^{1/4}) = o_p\{(\log p)^{-1/2}\}\), and the residual term \(R_{n, j}\) satisfies \(\max_{j\in[p]}|R_{n, j}| = o_p\{(\log p)^{-1/2}\}\).
Then, 
\begin{equation}\label{eq:step1-1}
    \sqrt{n}(\hat{\psi}_j - \psi_j)\hat{\beta}_j/2 = \tilde{\bG}_{j}\beta_j + Q_{n, j}\beta_j + R_{n, j}\beta_j + \sqrt{n}(\hat{\psi}_j - \psi_j)(\hat{\beta}_j - \beta_j)/2.        
\end{equation}
By Lemma~\ref{lem:exp}, we have 
\[
\max_{j\in[p]}|\sqrt{n}(\hat{\psi}_j - \psi_j)/2| \leq \max_{j\in[p]}|\tilde{\bG}_{j}| + o_p\{(\log p)^{-1/2}\} = O_p\{(\log p)^{1/2}\}.
\]
So, with Assumption~\ref{asmp:beta},
\[
    \max_{j\in[p]}|\sqrt{n}(\hat{\psi}_j - \psi_j)(\hat{\beta}_j - \beta_j)/2| \leq \max_{j\in[p]}|\sqrt{n}(\hat{\psi}_j - \psi_j)/2|\max_{j\in[p]}|\hat{\beta}_j - \beta_{j}| = o_p\{(\log p)^{-1/2}\}.
\]
Due to the finite upper bound of \(\beta_j\), \eqref{eq:step1} holds.

\Step{Step~2} (Large deviation approximation).\quad 
We show that
\begin{equation}\label{eq:step2}
    \max_{j\in\cH_0}\sup_{t\in(0, t^*]} \left|\frac{\P(\tilde{\bG}_{j}\beta_j \geq t)}{1-\Phi\{t/(\sigma_j\beta_j)\}} - 1\right| \leq C(\log p_0)^{-3/2}.
\end{equation}
Define \(\varpi = 2u_{\sigma, 0}^2u_{\beta, 0}^2/(\ell_{\sigma, 0}^2\ell_{\beta, 0}^2)\).
By Corollary~A.6 in the Supplementary Material of \cite{Dai2023} and \(\E[\max_{j\in[p]}|{\nu}_j(Z)|^{\theta}] < \infty\) for some constants \(\theta > 2 + r \varpi\) by Assumption~\ref{asmp:moment}, 
we have for any \(j\in\cH_0\), 
\[
    \sup_{t\in\left[0, \beta_{j}\sigma_j\sqrt{\varpi\log p_0}\right]}\left|\frac{\P(\tilde{\bG}_{j}\beta_{j} \geq t)}{1 - \Phi\{t/(\sigma_j\beta_{j})\}} - 1\right| \leq C(\log p_0)^{-3/2},
\]
where \(C > 0\) is a constant.
By \eqref{eq:G} and Assumption~\ref{asmp:Ssignal}, 
we have \(p_0G(t^*) \rightarrow\infty\), and \(p_0G\{\ell_{\beta, 0}\ell_{\sigma, 0}(\varpi\log p_0)^{1/2}\} \rightarrow 0\) by the definition of \(\varpi\).
Then, \(t^* < \ell_{\beta, 0}\ell_{\sigma, 0}\sqrt{\varpi \log p_0} \leq \beta_{j}\sigma_j\sqrt{\varpi\log p_0}\) for any \(j\in\cH_0\) and \eqref{eq:step2} holds.

\Step{Step~3} (Convergence of empirical measures).\quad We show that 
\begin{equation}\label{eq:step3}
    \sup_{t\in(0, t^*]}\left|\frac{\sum_{j\in \cH_0}\1(\tilde{\bG}_{j}\beta_{j} \geq t)}{p_0 G(t)} - 1\right| = o_p(1).
\end{equation}
Let \(b_p\) be a sequence satisfying \(b_p \rightarrow\infty\) and \(b_p \leq p_0/4\).
Let \(z_0 < z_1 < \cdots < z_{d_p} = 1/2\), \(t_i = G^{-1}(z_i)\), where \(z_0 = G(t^*)\), \(z_i = b_p/p_0 + b_p^{2/3}e^{i^\delta}/p_0\) for \(i\in[d_p]\), and \(d_p = [\log\{(p_0/2 - b_p)/b_p^{2/3}\}]^{1/\delta}\) for some \(0 < \delta < 1\). Note that \(G(t_i)/G(t_{i+1}) = 1+o(1)\) uniformly over \(i\) as \(p_0\rightarrow\infty\).
For any \(t\in[t_i, t_{i+1})\), 
\[\begin{split}
    &\frac{\sum_{j\in \cH_0} \1(\tilde{\bG}_{j}\beta_{j} \geq t)}{p_0 G(t)} \geq \frac{\sum_{j\in \cH_0} \1(\tilde{\bG}_{j}\beta_{j} \geq t_{i+1})}{p_0 G(t_{i+1})} \frac{G(t_{i+1})}{G(t_i)}\\ 
    &\frac{\sum_{j\in \cH_0} \1(\tilde{\bG}_{j}\beta_{j} \geq t)}{p_0 G(t)} \leq \frac{\sum_{j\in \cH_0} \1(\tilde{\bG}_{j}\beta_{j} \geq t_i)}{p_0 G(t_{i})} \frac{G(t_{i})}{G(t_{i+1})}.
\end{split}\]
It therefore suffices to show 
\begin{equation}\label{eq:step3-1}
    \max_{i\in\{0\}\cup[d_p]}\left|\frac{\sum_{j\in \cH_0} \1(\tilde{\bG}_{j}\beta_{j} \geq t_i)}{p_0 G(t_i)} - 1\right| \overset{p}{\rightarrow} 0.
\end{equation}
By the uniform convergence \eqref{eq:step2}, for \(\forall\epsilon > 0\), for large enough \(n, p\), we have 
\[\begin{split}
    &\P\left(\max_{i\in\{0\}\cup[d_p]}\left|\frac{\sum_{j\in \cH_0} \1(\tilde{\bG}_j\beta_{j} \geq t_i)}{p_0 G(t_i)} - 1\right| \geq \epsilon\right)\\ 
    &\qquad \leq \P\left(\max_{i\in\{0\}\cup[d_p]}\left|\frac{\sum_{j\in \cH_0} \1(\tilde{\bG}_j\beta_{j} \geq t_i) - \P(\tilde{\bG}_j\beta_{j} \geq t_i)}{p_0 G(t_i)}\right| \geq \epsilon/2\right)\\
    &\qquad \leq 4\epsilon^{-2}\sum_{i = 0}^{d_p} \frac{1}{p_0^2G^2(t_i)} \E\left[\sum_{j\in \cH_0} \left\{\1(\tilde{\bG}_j\beta_{j} \geq t_i) - \P(\tilde{\bG}_j\beta_{j} \geq t_i)\right\}\right]^2
\end{split}\]
Define $D(t) = \E\left[\sum_{j\in \cH_0} \left\{\1(\tilde{\bG}_j\beta_{j} \geq t) - \P(\tilde{\bG}_j\beta_{j} \geq t)\right\}\right]^2.$
By calculation, we have 
\[\begin{split}
    D(t) &= \sum_{j\in \cH_0} [\P(\tilde{\bG}_j\beta_{j} \geq t) - \{\P(\tilde{\bG}_j\beta_{j} \geq t)\}^2]\\
    &\qquad + \sum_{j, k\in \cH_0, j\neq k}\left\{\P(\tilde{\bG}_j\beta_{j} \geq t, \tilde{\bG}_k\beta_{k} \geq t) - \P(\tilde{\bG}_j\beta_{j} \geq t)\P(\tilde{\bG}_k\beta_{k} \geq t)\right\}\\
    &:= V_1(t) + V_2(t).
\end{split}\]
For \(V_1(t)\), by \eqref{eq:step2}, we have
\[
    V_1(t) \leq \sum_{j\in \cH_0} \P(\tilde{\bG}_j\beta_{j} \geq t) = p_0G(t)\{1+C(\log p_0)^{-3/2}\},
\]
uniformly for \(t \in (0, t^*]\).
By Lemma~A.19 in the Supplementary Material of \cite{Dai2023} and Assumption~\ref{asmp:moment}(b) 
and Assumption~\ref{asmp:cov}, 
we have, for \(j\neq k\in \cH_0\),
\[
    \sup_{t\in\left[0, \ell_{\sigma, 0}\ell_{\beta, 0}\sqrt{\varpi\log p_0}\right]}\left|\frac{\P(\tilde{\bG}_j\beta_{j} \geq t, \tilde{\bG}_k\beta_{k} \geq t)}{[1-\Phi\{t/(\sigma_j\beta_{j})\}][1-\Phi\{t/(\sigma_k\beta_{k})\}]} - 1\right| \leq C(\log p_0)^{-1-\gamma_1},
\] 
for some constant \(C > 0\) and \(\gamma_1 = \min(1/2, \gamma-2)\).
Together with \eqref{eq:step2} and \(t^* \leq \ell_{\beta, 0}\ell_{\sigma, 0}\sqrt{\varpi \log p_0}\), we have 
\[\begin{split}
    V_2(t) & \leq \sum_{j\neq k\in \cH_0} [1-\Phi\{t/(\sigma_j\beta_{j})\}][1-\Phi\{t/(\sigma_k\beta_k)\}] \left[\{1+C(\log p_0)^{-1-\gamma_1}\} - \{1 - C(\log p_0)^{-3/2}\}^2\right]\\
    &\leq 3C(\log p_0)^{-1-\gamma_1}\{p_0G(t)\}^2 
\end{split}\]
uniformly for \(t\in(0, t^*]\). Then, 
\[
    D(t) \leq p_0G(t)\{1+C(\log p_0)^{-3/2}\} + 3C(\log p_0)^{-1-\gamma_1}\{p_0G(t)\}^2
\] 
uniformly for \(t\in(0, t^*]\). Turning back to \eqref{eq:step3-1}, we have 
\[\begin{split}
    &\P\left(\max_{i\in\{0\}\cup[d_p]}\left|\frac{\sum_{j\in \cH_0} \1(\tilde{\bG}_j\beta_{j} \geq t_i)}{p_0 G(t_i)} - 1\right| \geq \epsilon\right)\\
    &\quad\leq 4\epsilon^{-2} \sum_{i = 0}^{d_p} \left\{\frac{1+C(\log p_0)^{-3/2}}{p_0G(t_i)} + 3C(\log p_0)^{-1-\gamma_1}\right\}\\
    &\quad\leq \frac{4C}{\epsilon^2 b_p^{2/3}}\sum_{i = 0}^{d_p} \frac{1}{b_p^{1/3} + e^{i^\delta}} + 12\epsilon^{-2}C (d_p+1) (\log p_0)^{-1-\gamma_1} \rightarrow 0, 
\end{split}\]
where the last inequality follows from \(b_p \rightarrow \infty\) as \(p\rightarrow\infty\). Thus, \eqref{eq:step3} follows.

\Step{Step~4} (Symmetry of agent-X importance under null).\quad 
We define four events 
\[\begin{split}
    \cE_1 &= \left\{\max_{j\in[p]} \left|Q_{n, j}\right| \leq \epsilon_p \right\},\qquad \cE_2 = \left\{\max_{j\in[p]} \left|R_{n, j}\right| \leq \epsilon_p\right\},\\
    \cE_3 &= \left\{\max_{j\in[p]}\left|\sqrt{n}(\hat{\psi}_{j} - \psi_j)(\hat{\beta}_{j} - \beta_{j})/2\right| \leq \epsilon_p\right\},\qquad \cE_4 = \left\{\sup_{t\in(0, t^*]}\left|\frac{\sum_{j\in \cH_0}\1(\tilde{\bG}_j\beta_{j} \geq t)}{p_0 G(t)} - 1\right| \leq \epsilon \right\},
\end{split}\]
where \(\epsilon_p = o\big(1/\sqrt{\log p}\big)\) is a nonrandom sequence.
We have \(\P(\bigcup_{\ell = 1}^4 \cE_\ell^c) \leq \sum_{\ell = 1}^4\P(\cE_\ell^c)\rightarrow 0\) as \(n, p\rightarrow\infty\).
By \(\bigcap_{\ell = 1}^4 \cE_\ell\), we have 
\[\begin{split}
    \frac{\sum_{j\in \cH_0} \1(\sqrt{n}\hat{\psi}_j\hat{\beta}_j/2 \geq t)}{p_0 G(t)} &\leq \frac{\sum_{j\in \cH_0} \1(\tilde{\bG}_j\beta_{j} \geq t-c_4\epsilon_p)}{p_0G(t)} \leq \frac{(1+\epsilon) G(t - c_4\epsilon_p)}{G(t)} \leq 1 + C \epsilon
\end{split}\]
holds uniformly over \(t\in (0, t^*]\), since \(G(t - o(1/\sqrt{\log p}))/G(t) = 1 + o(1)\) uniformly over \(t\in(0, t^*]\), where \(c_4 = 1+2u_{\beta, 0}\). 
Similarly, we have 
\[
    \frac{\sum_{j\in \cH_0} \1(\sqrt{n}\hat{\psi}_{j}\hat{\beta}_{j}/2 \geq t)}{p_0 G(t)} \geq 1 - C \epsilon.
\]
Then, 
\[
    \sup_{t\in(0, t^*]}\left|\frac{\sum_{j\in \cH_0}\1(\sqrt{n}\hat{\psi}_{j}\hat{\beta}_{j}/2 \geq t)}{p_0 G(t)} - 1\right| = o_p(1).
\]

\Step{Step~5} (FDR control at \(t^*\)).\quad 
The left is to show \(L \leq t^*\), which is equivalent to \(\widehat{\FDP}(t^*) \leq \alpha\), where \(\widehat{\FDP}(t)\) is the estimated FDR level for the threshold \(t\), defined as 
\[
    \widehat{\FDP}(t) = \frac{\sum_{j\in[p]} \1(\sqrt{n}\hat{\psi}_{j}\hat{\beta}_{j}/2 \leq -t)}{1\vee\sum_{j\in[p]}1(\sqrt{n}\hat{\psi}_{j}\hat{\beta}_{j}/2 \geq t)}.
\] 
Define three events
\[\begin{split}
    \cE_5 &= \left\{\min_{j\in\cH_1} \hat{\beta}_j > \ell_{\beta, 1}/2\right\},\qquad \cE_6 = \left\{\sum_{j\in\cH_1} \1(\sqrt{n}\hat{\psi}_j\hat{\beta}_j/2 \leq -t^*) = 0\right\},\\ 
    \cE_7 &= \left\{\sum_{j\in\cH_1} \1(\sqrt{n}\hat{\psi}_j\hat{\beta}_j/2 \geq t^*) \geq p_{1*}\right\}.
\end{split}\]
For \(\cE_5\), we have
\[
    \P(\cE_5^c) \leq \P\left(\min_{j\in \cH_1}\beta_j - \max_{j\in\cH_1}|\hat{\beta}_{j} - \beta_{j}| \leq \ell_{\beta, 1}/2\right) = \P\left(\max_{j\in \cH_1}|\hat{\beta}_{j} - \beta_{j}| \geq \ell_{\beta, 1}/2\right) \rightarrow 0
\]
by Assumption~\ref{asmp:beta}.
Then, for \(\cE_6\), 
\[\begin{split}
\P(\cE_6^c) &= \P\left\{\sum_{j\in\cH_1}\1(\sqrt{n}\hat{\psi}_j\hat{\beta}_j/2 \leq -t^*) \geq 1\right\}\\    
&\leq \P\left(\bigcup_{j\in\cH_1}\{\sqrt{n}\hat{\psi}_j\hat{\beta}_j/2 \leq -t^*\}, \cE_1\cap\cE_2\cap\cE_3\cap\cE_5\right) + \P\left(\cE_1^c\cup\cE_2^c\cup\cE_3^c\cup\cE_5^c\right).
\end{split}\]
Recalling the decomposition of \(\sqrt{n}(\hat{\psi}_j - \psi_j)\hat{\beta}_j/2\) in \eqref{eq:step1-1}, we have 
\[
    \sqrt{n}\hat{\psi}_{j}\hat{\beta}_{j} = \tilde{\bG}_j\beta_{j} + Q_{n, j}\beta_{j} + R_{n, j}\beta_{j} + \sqrt{n}(\hat{\psi}_{j} - \psi_j)(\hat{\beta}_{j} - \beta_{j})/2 + \sqrt{n}\psi_j\hat{\beta}_{j}/2.
\]
Then,
\[\begin{split}
    \P\left[\bigcup_{j\in \cH_1}\{\sqrt{n}\hat{\psi}_{j}\hat{\beta}_{j}/2 \leq -t^*\}, \cE_1\cap\cE_2\cap\cE_3\cap\cE_5\right] & \leq \sum_{j\in \cH_1}\P\left(\tilde{\bG}_j\beta_{j}\leq -t^* + c_5\epsilon_p - \sqrt{n}\psi_j\ell_{\beta, 1}/4\right)\\
    & \leq \sum_{j\in \cH_1}\P\left(\tilde{\bG}_j\beta_{j}\leq -t^* + c_5\epsilon_p\right),\\
\end{split}\]
where \(c_5 = 1+2u_{\beta, 1}\) is a constant.
Similar to Step~2, it can be shown that 
\[
    \max_{j\in\cH_1}\sup_{t \in (0, t^*]}\left|\frac{\P(\tilde{\bG}_j\beta_{j} \leq -t)}{1 - \Phi\{t/(\sigma_j\beta_{j})\}} - 1\right| = o(1).
\]
By Assumption~\ref{asmp:Ssignal}, 
we have \(G((\eta\log p_0)^{1/2})/G(t^*) \rightarrow \infty\). Then, \(t^* > \sqrt{\eta\log p_0}\).
For large enough \(n, p\), 
\[\begin{split}
    \P\left[\bigcup_{j\in \cH_1}\{\sqrt{n}\hat{\psi}_{j}\hat{\beta}_{j}/2 \leq -t^*\}, \cE_1\cap\cE_2\cap\cE_3\cap\cE_7\right] & \leq \sum_{j\in \cH_1}\P\left(\tilde{\bG}_j\beta_{j}\leq -t^* + c_5\epsilon_p\right)\\
    & \leq \sum_{j\in \cH_1}\P\left(\tilde{\bG}_j\beta_{j}\leq -\sqrt{(\eta/2)\log p_0}\right)\\
    & \leq (1+\epsilon)\sum_{j\in \cH_1}\left(1-\Phi\left[\{(\eta/2)\log p_0\}^{1/2}/(\sigma_j\beta_j)\right]\right)\\
    & \lesssim p_1(\log p_0)^{-1/2}p_0^{-\eta/(4u_{\sigma, 1}^2u_{\beta, 1}^2)} \rightarrow 0, 
\end{split}\]
where the last inequality holds due to \(1-\Phi(x)\leq \exp(-x^2/2)/(x\sqrt{2\pi})\).
So, \(\P(\cE_6^c) \rightarrow 0\) as \(n, p\rightarrow\infty\).

Next, we consider the event \(\cE_7\). We have 
\[\begin{split}
    \P(\cE_7^c) &= \P\left\{\sum_{j\in \cH_1}\1(\sqrt{n}\hat{\psi}_{j}\hat{\beta}_{j}/2 \geq t^*) < p_{1*}\right\} \leq \P\left(\bigcup_{j\in \cH_{1*}}\left\{\sqrt{n}\hat{\psi}_{j}\hat{\beta}_{j}/2 < t^*\right\}\right)\\
    &\leq \P\left[\bigcup_{j\in \cH_{1*}} \left\{\sqrt{n}(\psi_j - \hat{\psi}_{j})\hat{\beta}_{j}/2 > \sqrt{n}\psi_j\hat{\beta}_{j}/2 - t^*\right\}, \cE_1\cap\cE_2\cap\cE_3\cap\cE_5\right]\\
    &\quad + \P(\cE_1^c\cup\cE_2^c\cup\cE_3^c\cup\cE_5^c)\\
    &\leq \P\left(\bigcup_{j \in \cH_{1*}}\left\{\tilde{\bG}_j > \sigma_j\sqrt{C\log p}\right\}\right) + \P(\cE_1^c\cup\cE_2^c\cup\cE_3^c\cup\cE_5^c) = o(1),
\end{split}\]
where the last inequality holds by Assumption~\ref{asmp:Ssignal} 
and \(t^* < \ell_{\beta, 0}\ell_{\sigma, 0}\sqrt{\varpi \log p_0}\), and the last equality holds by Lemma~\ref{lem:exp}.

Define two events
\[\begin{split}
    \cE_8 &= \left\{\sup_{t\in(0, t^*]}\left|\frac{\sum_{j\in \cH_0}\1(\sqrt{n}\hat{\psi}_{j}\hat{\beta}_{j}/2 \leq -t)}{p_0G(t)} - 1\right| \leq \epsilon\right\},\\
    \cE_{9} &= \left\{\sup_{t\in(0, t^*]}\left|\frac{\sum_{j\in \cH_0}\1(\sqrt{n}\hat{\psi}_{j}\hat{\beta}_{j}/2 \geq t)}{p_0G(t)} - 1\right| \leq \epsilon\right\}.
\end{split}\] 
Similar to Step~3--4, we have \(\P(\cE_8^c) \rightarrow 0\) as \(n, p\rightarrow\infty\).
Then, by \(\bigcap_{\ell = 6}^9\cE_\ell\), 
\[\begin{split}
    \widehat{\FDP}(t^*) &\leq \frac{\sum_{j\in \cH_0} \1(\sqrt{n}\hat{\psi}_{j}\hat{\beta}_{j}/2 \leq -t^*)}{\sum_{j\in \cH_0}\1(\sqrt{n}\hat{\psi}_{j}\hat{\beta}_{j}/2 \geq t^*) + p_{1*}} \leq \frac{(1+\epsilon) G(t^*)}{(1 - \epsilon) G(t^*) + p_{1*}/p_0} = \frac{(1+\epsilon)\alpha}{2 - (1+\epsilon)\alpha}\leq \alpha.
\end{split}\]
Then, \(\P(\widehat{\FDP}(t^*)>\alpha)\leq \P(\bigcup_{\ell = 6}^9 \cE_\ell^c) \leq \sum_{j = 6}^9 \P(\cE_\ell^c)\rightarrow 0\), which implies \(\P(L \leq t^*)\rightarrow 1\) as \(n, p\rightarrow\infty\) by the definition of \(L\).
Finally, letting \(\cE_{10} = \{L\leq t^*\}\), we have by \(\bigcap_{\ell = 8}^{10} \cE_\ell\),
\[\begin{split}
    \FDP_\cA &= \frac{\sum_{j\in \cH_0}\1(\sqrt{n}\hat{\psi}_{j}\hat{\beta}_{j}/2 \geq L)}{\sum_{j\in[p]} \1(\sqrt{n} \hat{\psi}_{j}\hat{\beta}_{j}/2 \geq L)} \leq \frac{(1+\epsilon)^2\sum_{j\in \cH_0}\1(\sqrt{n}\hat{\psi}_{j}\hat{\beta}_{j}/2 \leq -L)}{\sum_{j\in[p]} \1(\sqrt{n} \hat{\psi}_{j}\hat{\beta}_{j}/2 \geq L)} \leq (1+3\epsilon)\alpha,
\end{split}\]
where the last inequality holds due to the definition of the threshold rule \eqref{eq:thrsh}.
We thus have \(\FDP_\cA \leq \alpha + o_p(1)\). Since for any \(\epsilon > 0\),
\[
    \FDR_{{\cA}} = \E\left[\FDP_\cA\right] \leq (1+3\epsilon)\alpha + \P\left\{\FDP_\cA > (1+3\epsilon)\alpha\right\}, 
\]
we have \({\lim\sup}_{n, p\rightarrow\infty} \FDR_{{\cA}} \leq \alpha\).

\subsection{Proof of other propositions}

  \begin{proof}[Proof of Proposition~\ref{prop:beta}]
    Proposition~\ref{prop:beta}(a) 
    naturally hold by \(\psi_j\geq 0\).
    For Proposition~\ref{prop:beta}(b), 
    recall that 
    \begin{equation*}
        \sqrt{n}(\hat{\beta}_j - \beta_{j}) = \frac{1}{\sqrt{n}}\sum_{i\in[n]} \nu_j(Z_i) + \tilde{R}_{n, j},
    \end{equation*}
    where \(\tilde{R}_{n, j}\) is the residual satisfying \(\max_{j\in[p]}|\tilde{R}_{n, j}| = O_p(\varrho_p)\).
    Then, we mainly analyze the term \(\max_{j\in[p]} |n^{-1/2}\sum_{i \in[n]} \nu_j(Z_i)|\).
    Let \(C_n = (nM\log n)^{1/\theta}\) for \(\theta > 2\). 
    We write 
    \[\begin{split}
        \nu_j(Z_i) &= \nu_j(Z_i)\1(|\nu_j(Z_i)|\leq C_n) - \E\left\{\nu_j(Z_i)\1(|\nu_j(Z_i)|\leq C_n)\right\}\\ 
        &\qquad  +  \nu_j(Z_i)\1(|\nu_j(Z_i)| > C_n) - \E\left\{\nu_j(Z_i)\1(|\nu_j(Z_i)| > C_n)\right\}\\
        &:= \nu_{ij, 1} + \nu_{ij, 2}.
    \end{split}\]
    Then, for some \(a \in (0, 1)\), 
    \[\begin{split}
        &\P\left\{\max_{j\in[p]}\left|\frac{1}{\sqrt{n}}\sum_{i = 1}^{n} \nu_j(Z_i)\right| \geq \sqrt{C\log p}\right\}\\ 
        &\qquad \leq \P\left\{\max_{j\in[p]}\left|\frac{1}{\sqrt{n}}\sum_{i = 1}^{n} \nu_{ij, 1}\right| \geq (1-a)\sqrt{C\log p}\right\} + \P\left\{\max_{j\in[p]}\left|\frac{1}{\sqrt{n}}\sum_{i = 1}^{n} \nu_{ij, 2}\right| \geq a\sqrt{C\log p}\right\}\\
        &\qquad := W_1 + W_2,
    \end{split}\]
    for some constant \(C > 2\max_{j\in[p]}\E\{\nu_j(Z)\}^2/(1-a)^2\).
    For \(W_1\), by Lemma~\ref{lem:Bernstein}, we have
    \[\begin{split}
        W_1 &\leq \sum_{j\in[p]}\P\left(\left|\frac{1}{\sqrt{n}}\sum_{i = 1}^n \nu_{ij, 1}\right| \geq (1-a)\sqrt{C\log p}\right)\\
        &\leq 2p\exp\left(-\frac{C\log p}{2\max_{j\in[p]}\E\{\nu_j(Z)\}^2/(1-a)^2 + \{(2\sqrt{C}/3)/(1-a)\}C_n\sqrt{\log p/n}}\right)\\
        &\leq 2p^{-1},
    \end{split}\]
    where the last inequality holds due to \(C_n\sqrt{\log p/n} \rightarrow 0\) as \(n, p\rightarrow\infty\).
    For \(W_2\), note that 
    \[\begin{split}
        &\sqrt{n}\max_{j\in[p]}\left|\E\{\nu_j(Z_i)\1(|\nu_j(Z_i)| > C_n)\}\right|\leq \max_{j\in[p]}\sqrt{n\E\{\nu_j(Z_i)\}^2 \P(|\nu_j(Z_i)| > C_n)}\\
        &\qquad \leq  \max_{j\in[p]}[\E\{\nu_j(Z_i)\}^2]^{1/2} \sqrt{\frac{n \max_{j\in[p]}\E|\nu_j(Z_i)|^\theta}{C_n^\theta}} \leq \max_{j\in[p]}[\E\{\nu_j(Z_i)\}^2]^{1/2}(\log n)^{-1/2}.
    \end{split}\]
    Then, for large \(n\), we have 
    \[\begin{split}
        W_2 &\leq \P\left(\frac{1}{\sqrt{n}}\sum_{i = 1}^n \max_{j\in[p]}|\nu_j(Z_i)|\1(|\nu_j(Z_i)| > C_n) \geq (a/2)\sqrt{C\log p}\right)\\
        &\leq \P\left(\max_{j\in[p]}|\nu_j(Z_i)| > C_n \text{ for some } i\in[n]\right)\\
        &\leq \sum_{i = 1}^n \P\left(\max_{j\in[p]}|\nu_j(Z_i)| > C_n\right) \leq \frac{n\E[\max_{j\in[p]}|\nu_j(Z)|^\theta]}{C_n^\theta} \leq (\log n)^{-1}.
    \end{split}\]
    Thus, 
    \[
        \P\left\{\max_{j\in[p]}\left|\frac{1}{\sqrt{n}}\sum_{i = 1}^{n} \nu_j(Z_i)\right| \geq \sqrt{C\log p}\right\} \leq 2p^{-1} + (\log n)^{-1},
    \]
    which means \(\max_{j\in [p]} |n^{-1/2}\sum_{i = 1}^n \nu_j(Z_i)| = O_p\big(\sqrt{\log p}\big)\). 
    Thus, 
    \[
        \max_{j\in[p]}|\hat{\beta}_j - \beta_{j}| = O_p\left\{(\sqrt{\log p} \vee \varrho_p)/\sqrt{n}\right\} = o_p\{(\log p)^{-1}\}.
    \]
  \end{proof}

  \begin{proof}[Proof of of uniform bounds of \(\beta_j\) for \ref{M1}--\ref{M3}]
    For \(\pred\) chosen from \ref{M1}--\ref{M3} and \(\beta_j = \pred(f^*, P_{Z^j})\), 
    \[
      \pred(f^*, P_Z) \leq \beta_j \leq \pred(f^*, P_Z) + u_{\beta},
    \]
    where
  \begin{itemize}
    \item[(i)] \( u_{\beta}=\max_{j\in \cH_1}\E\{f^*(X) - f^*(X_{-j}, \tilde{X}_j)\}^2\) for \ref{M1};
    \item[(ii)] \(u_{\beta}= \max_{j\in \cH_1}\E(f^*(X)\log\{f^*(X)/f^*(X_{-j}, \tilde{X}_j)\} + (1-f^*(X))\log[\{1-f^*(X)\}/\{1-f^*(X_{-j}, \tilde{X}_j)\}])\) for \ref{M2};
    \item[(iii)] \(u_{\beta}= \max_{j\in \cH_1}\E[f^*(X)\log\{f^*(X)/f^*(X_{-j}, \tilde{X}_j)\} - \{f^*(X) - f^*(X_{-j}, \tilde{X}_j)\}]\) for \ref{M3}. 
  \end{itemize}
    
  For Assumption~\ref{asmp:beta}(a), the lower bound of \(\beta_j\), \(\pred(f^*, P_Z)\), is usually greater than zero when noise exists, while the upper bound is finite under some mild conditions on \(f^*\), such as \(f^*\in\ell^\infty(\bR^p)\) for \(\pred\) in \ref{M1}, 
  where \(\ell^\infty(\bR^p)\) is the set of uniformly bounded real functions on \(\bR^p\). 
\end{proof}

\begin{proof}[Proof of Proposition~\ref{prop:power}]
    For notational convenience, we rewrite the SNR as 
    \[
    \SNR(\hat{\psi}_j^\B; \cH_1, \cH_0) = \frac{|\cH_1|^{-1}\sum_{j\in\cH_1} \E\{n(\hat{\psi}_j^\B)^2/4\}}{|\cH_0|^{-1}\sum_{j\in\cH_0} \E\{n(\hat{\psi}_j^\B)^2/4\}}.
    \]
    To prove the conclusion, it suffices to show that 
    \[\begin{split}
        &\frac{\SNR(\hat{\psi}_j; \cH_1, \cH_0)}{|\cH_1|^{-1}\sum_{j\in\cH_1}(n\psi_j^2/4 + \sigma_j^2)/(|\cH_0^{-1}|\sum_{j\in\cH_0}\sigma_j^2)} \rightarrow 1,\\ 
        &\frac{\SNR(\hat{\psi}_j^\B; \cH_1, \cH_0)}{|\cH_1|^{-1}\sum_{j\in\cH_1}\beta_j^2(n\psi_j^2/4 + \sigma_j^2)/\{\pred^2(f^*, P_Z)|\cH_0^{-1}|\sum_{j\in\cH_0}\sigma_j^2\}}\rightarrow 1.
    \end{split}\]
    
    Recall that
$\frac{\sqrt{n}}{2}\left(\hat{\psi}_{j} - \psi_{j}\right) = \tilde{\bG}_j + Q_{n, j} + R_{n, j}$ where \(\E(\tilde{\bG}_j^2) = [\E\{\nu_0(Z)\}^2 + \E\{\nu_j(Z)\}^2]/2 = \sigma_j^2\) and $\E\{R_{n,j}^2\}=o(1)$. 
From \eqref{eq:supp:Unj} in the proof of Lemma~\ref{supp:lem:removeDep}, we have
    \begin{equation}
        \begin{split}
            \E\{Q_{n, j}^2\} &\leq \E\left(n^{-1}\sum_{i\in\cJ_{n, j}}\E[\{\nu_0(Z_i) + \nu_{j}(Z_i)\}^2\mid \cJ_{n, j}]\right)\\
            &\leq 4\sigma_j^2\E|\cJ_{n, j}|/n \\
            &\leq 2\sigma_j^2\{\sqrt{\log 2} + \sqrt{\pi}/2\}\sqrt{2/n},
        \end{split}\label{eq:EU}
    \end{equation}
    where the last inequality holds by \eqref{eq:J} and Lemma~\ref{lem:exp-Hoeffding}.
    Thus, we get 
    \[\begin{split}
        \E[\{\sqrt{n}(\hat{\psi}_j - \psi_j)/2\}^2] &= \E\{(\tilde{\bG}_j + Q_{n, j} + R_{n, j})^2\}\\
        &= \E\{\tilde{\bG}_j^2\} + \E\{(Q_{n, j} + R_{n, j})^2\} + 2\E\{\tilde{\bG}_j(Q_{n, j} + R_{n, j})\}\\
        &= \sigma_j^2 + o(1),
    \end{split}\]
    where the interactions are controlled using Cauchy-Schwarz inequality. 

    For \(j\in\cH_0\) (i.e., \(\psi_j = 0\)), by \eqref{eq:EU} and \(\max_{j\in\cH_0}\E\{R_{n, j}^2\} = o(1)\), we have \(\max_{j\in\cH_0}|\E(n\hat{\psi}_j^2/4) - \sigma_j^2| = o(1)\), and thus 
    \[
    \frac{|\cH_0|^{-1}\sum_{j\in\cH_0}\E(n\hat{\psi}_j^2/4)}{|\cH_0|^{-1}\sum_{j\in\cH_0}\sigma_j^2} = 1 + o(1).
    \]
    For \(j\in\cH_1\), we make the following decomposition, 
    \[\begin{split}
        \E\{n\hat{\psi}_j^2/4\} &= \E[\{\sqrt{n}(\hat{\psi}_j - \psi_j)/2\}^2] + n\psi_j^2/4 + \sqrt{n}\psi_j\E\{\sqrt{n}(\hat{\psi}_j - \psi_j)/2\}\\
        &= \E[\{\sqrt{n}(\hat{\psi}_j - \psi_j)/2\}^2] + n\psi_j^2/4 +  \sqrt{n}\psi_j \E(U_{n, j} + R_{n, j}).
    \end{split}\] 
    Then, it is easy to derive that
    \[
        \max_{j\in\cH_1}|\E\{n\hat{\psi}_j^2/4\} - \sigma_j^2 - n\psi_j^2/4|/(\sqrt{n}\psi_j) = o(1),
    \]
    and thus 
    \[
        \frac{|\cH_1|^{-1}\sum_{j\in\cH_1}\E(n\hat{\psi}_j^2/4)}{|\cH_1|^{-1}\sum_{j\in\cH_1}(n\psi_j^2/4 + \sigma_j^2)} = 1+ o(1).
    \]
    So, 
    \[\begin{split}
        &\frac{\SNR(\hat{\psi}_j; \cH_1, \cH_0)}{|\cH_1|^{-1}\sum_{j\in\cH_1}(n\psi_j^2/4 + \sigma_j^2)/(|\cH_0^{-1}|\sum_{j\in\cH_0}\sigma_j^2)}\\ 
        &\qquad =  \frac{|\cH_1|^{-1}\sum_{j\in\cH_1}\E(n\hat{\psi}_j^2/4)}{|\cH_1|^{-1}\sum_{j\in\cH_1}(n\psi_j^2/4 + \sigma_j^2)} \times \frac{|\cH_0|^{-1}\sum_{j\in\cH_0}\sigma_j^2}{|\cH_0|^{-1}\sum_{j\in\cH_0}\E(n\hat{\psi}_j^2/4)} = 1+ o(1).
    \end{split}\]
    
    Similarly, using the asymptotic expression \eqref{eq:beta-expr} 
    of \(\hat{\beta}_j\), \(\max_{j\in\cH_1}\psi_j \leq C\) and \(\max_{j\in[p]}\E(\tilde{R}^4_{n, j})\vee\E(R^4_{n, j}) = o(1)\), it can be derived that 
    \[
    \max_{j\in[p]}\left|\E[\{\sqrt{n}\hat{\psi}_j\hat{\beta}_j/2\}^2] - \E[\{\sqrt{n}\hat{\psi}_j{\beta}_j/2\}^2]\right| = o(1), 
    \]
    and further
    \[
    \frac{|\cH_1|^{-1}\sum_{j\in\cH_1}\E\{n(\hat{\psi}_j^\B)^2/4\}}{|\cH_1|^{-1}\sum_{j\in\cH_1}\beta_j^2(n\hat{\psi}_j^2/4 + \sigma_j^2)} = 1+o(1), \quad \frac{|\cH_0|^{-1}\sum_{j\in\cH_0}\E\{n(\hat{\psi}_j^\B)^2/4\}}{|\cH_0|^{-1}\sum_{j\in\cH_0}\beta_j^2\sigma_j^2} = 1+o(1).
    \]
    Due to \(\beta_j = \pred(f^*, P_Z)\) for all \(j\in\cH_0\), we have
    \[\begin{split}
        &\frac{\SNR(\hat{\psi}_j^\B; \cH_1, \cH_0)}{|\cH_1|^{-1}\sum_{j\in\cH_1}\beta_j^2(n\psi_j^2/4 + \sigma_j^2)/\{\pred^2(f^*, P_Z)|\cH_0^{-1}|\sum_{j\in\cH_0}\sigma_j^2\}}\\ 
        &\quad =  \frac{|\cH_1|^{-1}\sum_{j\in\cH_1}\E\{n(\hat{\psi}_j^E)^2/4\}}{|\cH_1|^{-1}\sum_{j\in\cH_1}\beta_j^2(n\psi_j^2/4 + \sigma_j^2)} \times \frac{\pred^2(f^*, P_Z)|\cH_0|^{-1}\sum_{j\in\cH_0}\sigma_j^2}{|\cH_0|^{-1}\sum_{j\in\cH_0}\E(n(\hat{\psi}_j^E)^2/4)} = 1+ o(1).
    \end{split}\]
    Thus, 
    \[
        \frac{\SNR(\hat{\psi}_j^\B; \cH_1, \cH_0)}{\SNR(\hat{\psi}_j; \cH_1, \cH_0)}\times\frac{\sum_{j\in\cH_1}(n\psi_j^2/4 + \sigma_j^2)\pred^2(f^*, P_Z)}{\sum_{j\in\cH_1}(n\psi_j^2/4 + \sigma_j^2)\beta_j^2} \rightarrow 1.
    \]  
\end{proof}

\subsection{Basic lemmas}

\begin{lemma}[Bernstein's inequality]\label{lem:Bernstein}
    Let \(X_1, \ldots, X_n\) be independent centered random variables a.s. bounded by \(A < \infty\) in absolute value. Let \(\sigma^2 = n^{-1}\sum_{i = 1}^n \E(X_i^2)\). Then, for all \(x >0\),
    \[
        \P\left(\sum_{i = 1}^n X_i \geq x\right) \leq \exp\left(-\frac{x^2}{2n\sigma^2 + 2A x/3}\right).
    \]
\end{lemma}

\begin{lemma}\label{lem:binomialization}
    Let \(N_n\sim \mathrm{Binomial(n, p)}\) and suppose when \(N = k\), \((\varepsilon_1, \ldots, \varepsilon_n)\) satisfies 
    \[
        \P\left(\varepsilon_1 = x_1, \ldots, \varepsilon_n = x_n\mid N = k\right) = \left\{\begin{array}{ll}
        1/\binom{n}{k}, & \mathrm{if}\ \sum_{i = 1}^n x_i = k,\\
        0, & \mathrm{otherwise,}
        \end{array}\right.
    \]
    where \((x_1, \ldots, x_n) \in \{1, 0\}^n\). Then, marginally, \(\varepsilon_1, \ldots, \varepsilon_n\) are independent Bernoulli, with \(\varepsilon_i\sim\mathrm{Bernoulli}(p)\).
\end{lemma}

  \begin{proof}[Proof of Lemma~\ref{lem:binomialization}]
    We verify the conclusion by definition.
    \[\begin{split}
        &\P(\varepsilon_1 = x_1, \ldots, \varepsilon_n = x_n)\\
        &\qquad = \sum_{k = 0}^n \P(\varepsilon_1 = x_1, \ldots, \varepsilon_n = x_n\mid N = k)\P(N = k)\\
        &\qquad = \P\left(\varepsilon_1 = x_1, \ldots, \varepsilon_n = x_n\mid N = \sum_{i = 1}^n x_i\right)\P\left(N = \sum_{i = 1}^n x_i\right)\\
        &\qquad = p^{\sum_{i = 1}^n x_i}(1 - p)^{n - \sum_{i = 1}^n x_i}\\
        &\qquad = \prod_{i = 1}^n p^{x_i}(1-p)^{1-x_i}. 
    \end{split}\]
    The conclusion follows.    
  \end{proof}

\begin{lemma}\label{lem:exp-Hoeffding}
    Let \(Z\) be a nonnegative random variable satisfying 
    $$\P(Z \geq t) \leq C\exp\left(-{t^2}/{v^2}\right),$$ 
    where \(C \geq 1\) and \(v > 0\) are constants. Then, 
    \[
        \E(Z) \leq v\left(\sqrt{\log C} + \sqrt{\pi}/2\right).
    \] 
\end{lemma}

  \begin{proof}[Proof of Lemma~\ref{lem:exp-Hoeffding}]
    For the nonnegative random variable \(Z\), we can write 
    \[\begin{split}
        \E(Z) &= \int_0^\infty \P(Z\geq t) \d t = \int_0^\infty 1\wedge C\exp(-t^2/v^2) \d t\\
        &\leq \int_0^{t_1} 1\d t + \int_{t_1}^\infty C\exp(-t^2/v^2) \d t = t_1 + \int_{t_1}^\infty C\exp(-t^2/v^2)\d t,
    \end{split}\]
    where \(t_1 = v \sqrt{\log C}\).
    Let \(t = v x\). Then, 
    \[
        \int_{t_1}^\infty C\exp(-t^2/v^2)\d t = v \int_{\sqrt{\log C}}^\infty C \exp(-x^2) \d x.
    \]
    Letting \(y = x - \sqrt{\log C}\), we have 
    \[
        x^2 - y^2 = \left(y + \sqrt{\log C}\right)^2 - y^2 = \log C + 2y\sqrt{\log C} \geq \log C, \quad y \geq 0.
    \]
    Equivalently, this implies \(C\exp(-x^2) \leq \exp(-y^2)\) for all \(x \geq \sqrt{\log C}\). Then, 
    \[
        \int_{\sqrt{\log C}}^\infty C \exp(-x^2)\d x \leq \int_{0}^\infty \exp(-y^2)\d y = \frac{\sqrt{\pi}}{2}.
    \]
    Putting them together, we obtain 
    \[
        \E(Z)\leq v\left(\sqrt{\log C} + \sqrt{\pi}/2\right).
    \]    
  \end{proof}

\section{Additional Simulations}\label{supp:simul}

\subsection{Implementation details}

We implement the Knockoff~\cite{Candes2018} method with \texttt{R} package \texttt{knockoff}. The code of BHq~\cite{Javanmard2019fdr} in \ref{S1}--\ref{S2} 
follows \url{https://web.stanford.edu/~montanar/sslasso/}.
The BHq method in real data analysis follows \url{https://web.stanford.edu/group/candes/knockoffs/software/knockoffs/tutorial-4-r.html}.
The data splitting~\cite{Dai2022ds} adopts the single split in~\ref{S1}--\ref{S2} 
and multiple splits in real data analysis, following the code of \url{https://github.com/Jeremy690/False-Discovery-Rate-via-Data-Splitting}. All methods which adopt the p-value-free threshold rule (including MAS, Knockoff, DS) use the following modified threshold to their statistics \(\{W_j: j\in[p]\}\) 
\[
    L = \inf\left\{t>0: \frac{1+\sum_{j\in[p]}\1(W_j\leq -t)}{1\vee    \sum_{j\in[p]}\1(W_j\geq t)}\leq \alpha\right\}.
\]

\subsection{Ablation experiments}

\begin{figure}[htbp]
  \centering
  \includegraphics[scale=0.6]{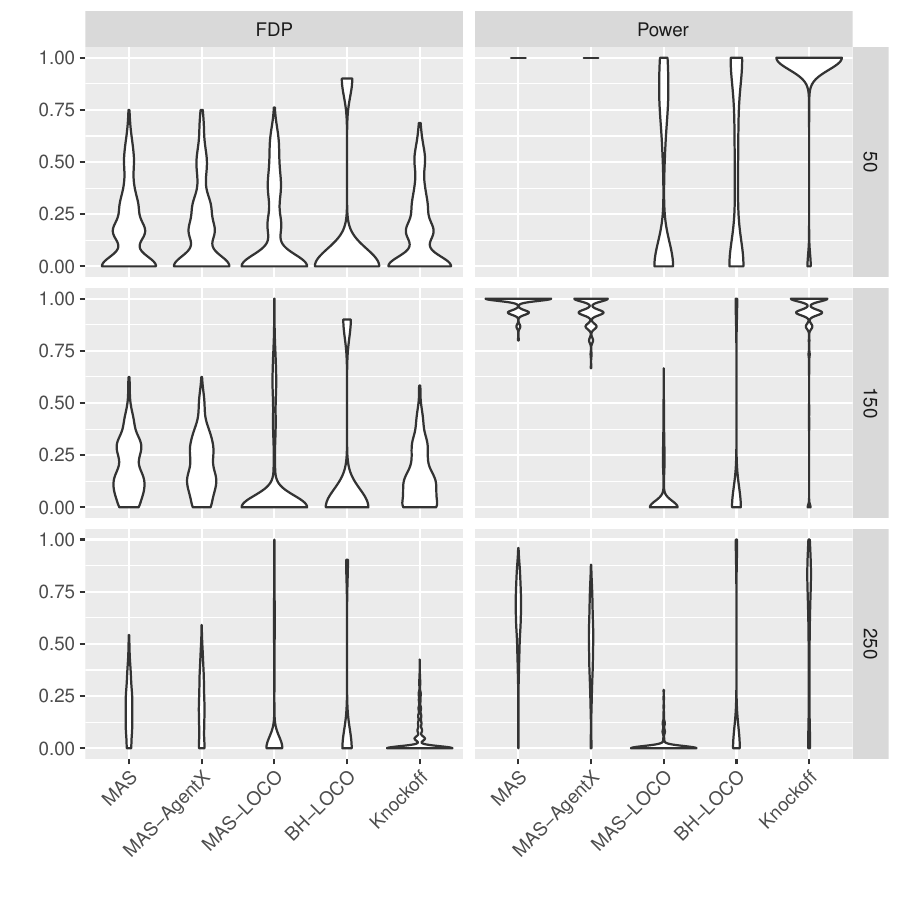}
  \caption{Violin plots of FDP and Power for each \(p\in\{50, 150, 250\}\) under the same setting~\ref{S1} as in Table~\ref{tab:02}}
  \label{fig:supp-tab02}
\end{figure}

We mentioned in the main text that strong correlations exist between p-values by BH-LOCO, which yields either no discovery or overwhelming discoveries. We provide the violin plots in Figure~\ref{fig:supp-tab02} for each \(p\in\{50, 150, 250\}\) under the same setting~\ref{S1} 
as in Table~\ref{tab:02}.
According to Table~\ref{tab:02}, 
all methods control the FDR. However, when we plot the FDP, such extreme performances of BH-LOCO are exposed. 
In the meantime, the methods based on MAS framework (with resplit-per-covariate strategy) present more reasonable FDP and power.

\subsection{Classification task}

In this section, we consider a classification task adopting the predictiveness measure function \(\pred\) in \ref{M2}. 
We consider the following logistic model:

\begin{enumerate}[label={(S\arabic*)}]  
    \setcounter{enumi}{3}
    \itemindent10pt
    \item \(Y\sim \mathrm{Bernoulli}(f(X^\T\eta))\), where \(f(z) = 1/\{1+\exp(-z)\}\), \(X\sim\mathcal{N}(0, (0.3^{\1(i\neq j)})_{p\times p})\), and \(\eta = (b_1\eta_1, \ldots, b_p\eta_p)^\T\) with \(\eta_j\sim\mathrm{Unif}([3, 5])\) and \(b_j\sim\mathrm{Unif}(\{-1, 1\})\) for \(j\in\cH_1\) and \(\eta_j = b_j = 0\) otherwise.\label{S4}
\end{enumerate}
The signal set \(\cH_1\) with size \(p_1 = 20\) are randomly chosen from \([p]\). 
We consider three popular benchmarks, including the BH procedure (BHq)~\cite{Ma2020}, the data splitting (DS) method~\cite{Dai2023}, the Model-X knockoffs with the approximate second-order Gaussian knockoffs and the difference of the regularization parameters as statistics (Knockoff)~\cite{Candes2018}. The learning algorithm \(\sL\) is specified as the Lasso.

\begin{figure}[htbp]
  \centering
  \includegraphics[scale=0.55]{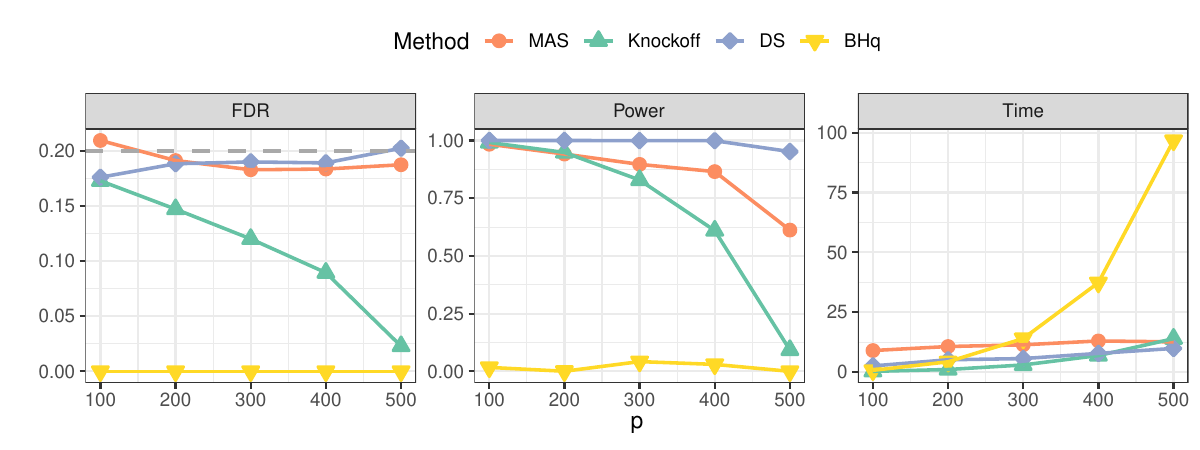}
  \caption{Comparison results of MAS, Knockoff~\cite{Candes2018}, DS~\cite{Dai2023} and BHq~\cite{Ma2020} under FDR level \(\alpha=0.2\) (darkgrey, dashed) in setting~\ref{S4}. The unit of Time is seconds}
  \label{fig:S3}
\end{figure}

Figure~\ref{fig:S3} reports the results under the logistic model~\ref{S4} when \(n = 600\) and \(p\in\{100, 200, \ldots, 500\}\). We can see that the FDR of all methods are controlled at the target level. 
Except that the strong dependence between covariates makes BHq extremely conservative, conclusions are similar to those for~\ref{S1}. 

\begin{thebibliography}{99}
    
\bibitem{Barber2015knockoff}
Barber R~F, Cand\`es E~J.
 Controlling the false discovery rate via knockoffs.
 Ann Statist, 2015, 43: 2055-2085

\bibitem{bates2021metropolized}
Bates S, Cand\`es E, Janson L, et~al.
 Metropolized knockoff sampling.
 J Amer Statist Assoc, 2021, 116: 1413-1427

\bibitem{Benjamini1995FDR}
Benjamini Y, Hochberg Y.
 Controlling the false discovery rate: a practical and powerful approach to multiple testing.
 J R Stat Soc Ser B Stat Methodol, 1995, 57: 289-300

\bibitem{Berrett2019CPT}
Berrett T~B, Wang Y, Barber R~F, et~al.
 The conditional permutation test for independence while controlling for confounders.
 J R Stat Soc Ser B Stat Methodol, 2020, 82: 175-197

\bibitem{GPT3}
Brown T, Mann B, Ryder N, et~al.
 Language models are few-shot learners.
 In: Proceedings of the 34th Annual Conference on Neural Information Processing Systems.
 Virtual-only, 2020, 1877-1901

\bibitem{Cai2022a}
Cai T~T, Sun W, Xia Y.
 L{AWS}: a locally adaptive weighting and screening approach to spatial multiple testing.
 J Amer Statist Assoc, 2022, 117: 1370-1383

\bibitem{Cai2024pGMC}
Cai L, Guo X, Zhong W.
 Test and measure for partial mean dependence based on machine learning methods.
 J Amer Statist Assoc, 2025, 120: 833-845

\bibitem{Candes2018}
Cand\`es E, Fan Y, Janson L, et~al.
 Panning for gold: `model-{X}' knockoffs for high dimensional controlled variable selection.
 J R Stat Soc Ser B Stat Methodol, 2018, 80: 551-577

\bibitem{Chen2023}
Chen G, Jia Y, Wang G, et~al.
 Zipper: addressing degeneracy in algorithm-agnostic inference.
 In: Proceedings of the 38th Annual Conference on Neural Information Processing Systems.
 Vancouver, 2024, 65884-65911

\bibitem{chernozhukov2018double}
Chernozhukov V, Chetverikov D, Demirer M, et~al.
 Double/debiased machine learning for treatment and structural parameters.
 Econom J, 2018, 21: C1-C68

\bibitem{Dai2022ds}
Dai C, Lin B, Xing X, et~al.
 False discovery rate control via data splitting.
 J Amer Statist Assoc, 2023, 118: 2503-2520

\bibitem{Dai2023}
Dai C, Lin B, Xing X, et~al.
 A scale-free approach for false discovery rate control in generalized linear models.
 J Amer Statist Assoc, 2023, 118: 1551-1565

\bibitem{Dai2024vim}
Dai B, Shen X, Pan W.
 Significance tests of feature relevance for a black-box learner.
 IEEE Trans Neural Netw Learn Syst, 2024, 35: 1898-1911

\bibitem{Du2021SDA}
Du L, Guo X, Sun W, et~al.
 False discovery rate control under general dependence by symmetrized data aggregation.
 J Amer Statist Assoc, 2023, 118: 607-621

\bibitem{LLM4Science}
Hayes T, Rao R, Akin H, et~al.
 Simulating 500 million years of evolution with a language model.
 Science, 2025, 387: 850-858

\bibitem{Hu2010gbh}
Hu J~X, Zhao H, Zhou H~H.
 False discovery rate control with groups.
 J Amer Statist Assoc, 2010, 105: 1215-1227

\bibitem{Javanmard2014}
Javanmard A, Montanari A.
 Confidence intervals and hypothesis testing for high-dimensional regression.
 J Mach Learn Res, 2014, 15: 2869-2909

\bibitem{Javanmard2019fdr}
Javanmard A, Javadi H.
 False discovery rate control via debiased lasso.
 Electron J Stat, 2019, 13: 1212-1253

\bibitem{DNN}
LeCun Y, Bengio Y, Hinton G.
 Deep learning.
 Nature, 2015, 521: 436-444

\bibitem{Lei2018loco}
Lei J, G'Sell M, Rinaldo A, et~al.
 Distribution-free predictive inference for regression.
 J Amer Statist Assoc, 2018, 113: 1094-1111

\bibitem{Li2018SABHA}
Li A, Barber R~F.
 Multiple testing with the structure-adaptive {B}enjamini-{H}ochberg algorithm.
 J R Stat Soc Ser B Stat Methodol, 2019, 81: 45-74

\bibitem{Li2021ggmknockoff}
Li J, Maathuis M~H.
 G{GM} knockoff filter: false discovery rate control for {G}aussian graphical models.
 J R Stat Soc Ser B Stat Methodol, 2021, 83: 534-558

\bibitem{Liu2022screenknockoff}
Liu W, Ke Y, Liu J, et~al.
 Model-free feature screening and {FDR} control with knockoff features.
 J Amer Statist Assoc, 2022, 117: 428-443

\bibitem{lu2018deeppink}
Lu Y, Fan Y, Lv J, et~al.
 {D}eep{PINK}: reproducible feature selection in deep neural networks.
 In: Proceedings of the 32nd Annual Conference on Neural Information Processing Systems.
 Montr\'eal, 2018, 

\bibitem{Lundborg2022vim}
Lundborg A~R, Kim I, Shah R~D, et~al.
 The projected covariance measure for assumption-lean variable significance testing.
 Ann Statist, 2024, 52: 2851-2878

\bibitem{Ma2020}
Ma R, Cai T~T, Li H.
 Global and simultaneous hypothesis testing for high-dimensional logistic regression models.
 J Amer Statist Assoc, 2021, 116: 984-998

\bibitem{Mi2021vim}
Mi X, Zou B, Zou F, et~al.
 Permutation-based identification of important biomarkers for complex diseases via machine learning models.
 Nat Commun, 2021, 12: 3008

\bibitem{Ning2017}
Ning Y, Liu H.
 A general theory of hypothesis tests and confidence regions for sparse high dimensional models.
 Ann Statist, 2017, 45: 158-195

\bibitem{DNN4Science}
Raissi M, Perdikaris P, Karniadakis G.
 Physics-informed neural networks: a deep learning framework for solving forward and inverse problems involving nonlinear partial differential equations.
 J Comput Phys, 2019, 378: 686-707

\bibitem{HIV2006}
Rhee S~Y, Taylor J, Wadhera G, et~al.
 Genotypic predictors of human immunodeficiency virus type 1 drug resistance.
 Proc Natl Acad Sci USA, 2006, 103: 17355-17360

\bibitem{Alessandro2019loco}
Rinaldo A, Wasserman L, G'Sell M.
 Bootstrapping and sample splitting for high-dimensional, assumption-lean inference.
 Ann Statist, 2019, 47: 3438-3469

\bibitem{romano2020deep}
Romano Y, Sesia M, Cand\`es E.
 Deep knockoffs.
 J Amer Statist Assoc, 2020, 115: 1861-1872

\bibitem{Scheidegger2022wGCM}
Scheidegger C, H\"{o}rrmann J, B\"{u}hlmann P.
 The weighted generalised covariance measure.
 J Mach Learn Res, 2022, 23: PaperNo.[273],68

\bibitem{Shah2020vim}
Shah R~D, Peters J.
 The hardness of conditional independence testing and the generalised covariance measure.
 Ann Statist, 2020, 48: 1514-1538

\bibitem{Srinivasan2021compoknockoff}
Srinivasan A, Xue L, Zhan X.
 Compositional knockoff filter for high-dimensional regression analysis of microbiome data.
 Biometrics, 2021, 77: 984-995

\bibitem{Geer2014}
van~de Geer S, B\"{u}hlmann P, Ritov Y, et~al.
 On asymptotically optimal confidence regions and tests for high-dimensional models.
 Ann Statist, 2014, 42: 1166-1202

\bibitem{Vaart1996}
van~der Vaart A~W, Wellner J~A.
 Weak Convergence and Empirical Processes: With Applications to Statistics.
 Springer New York, 1996

\bibitem{Verdinelli2024decorrelated}
Verdinelli I, Wasserman L.
 Decorrelated variable importance.
 J Mach Learn Res, 2024, 25: PaperNo.[7],27

\bibitem{Williamson2021vim}
Williamson B~D, Gilbert P~B, Carone M, et~al.
 Nonparametric variable importance assessment using machine learning techniques.
 Biometrics, 2021, 77: 9-22

\bibitem{Williamson2023vim}
Williamson B~D, Gilbert P~B, Simon N~R, et~al.
 A general framework for inference on algorithm-agnostic variable importance.
 J Amer Statist Assoc, 2023, 118: 1645-1658

\bibitem{WOLPERT1992stacking}
Wolpert D~H.
 Stacked generalization.
 Neural Netw, 1992, 5: 241-259

\bibitem{Wood2024vim}
Wood D, Papamarkou T, Benatan M, et~al.
 Model-agnostic variable importance for predictive uncertainty: an entropy-based approach.
 Data Min Knowl Discov, 2024, 38: 4184-4216

\bibitem{Zhang2013}
Zhang C~H, Zhang S~S.
 Confidence intervals for low dimensional parameters in high dimensional linear models.
 J R Stat Soc Ser B Stat Methodol, 2014, 76: 217-242

\end{thebibliography}
\end{document}